\documentclass[12pt]{article}
\usepackage{hyperref}
\usepackage{amsmath, amsfonts, amssymb, mathrsfs}
\usepackage{dcolumn}
\usepackage{caption}
\usepackage{subcaption}
\usepackage{filemod}
\usepackage[ruled, vlined]{algorithm2e}
\usepackage{floatrow}
\usepackage{graphicx}
\usepackage{setspace}
\usepackage[title]{appendix}
\graphicspath{{./Figures/}}

\usepackage{natbib}

\title{\vspace{-1cm}Traditional kriging versus modern Gaussian processes\\
for large-scale mining data}
\author{Ryan B. Christianson\thanks{Corresponding author: {\tt 
rchristianson@vt.edu}, Department of Statistics, Virginia Tech} \and 
Ryan M. Pollyea\thanks{Department of Geosciences, Virginia Tech 926 W. Campus 
Dr. Blacksburg, VA 24060} \and Robert B. Gramacy\thanks{Department of 
Statistics, Virginia Tech, 250 Drillfield Dr. Blacksburg, VA 24061}}
\date{\today}

\begin{document}
\maketitle

\vspace{-0.5cm}

\begin{abstract}
The canonical technique for nonlinear modeling of spatial/point-referenced
data is known as kriging in geostatistics, and as Gaussian Process (GP)
regression for surrogate modeling and statistical learning. This article reviews
many similarities shared between kriging and GPs, but also highlights some
important differences.  One is that GPs impose a process that can be used to
automate kernel/variogram inference, thus removing the human from the loop.
The GP framework also suggests a probabilistically valid means of scaling to
handle a large corpus of training data, i.e., an alternative to so-called
ordinary kriging. Finally, recent GP implementations are tailored to make the
most of modern computing architectures such as multi-core workstations and
multi-node supercomputers. We argue that such distinctions are important even
in classically geostatistical settings.  To back that up, we present
out-of-sample validation exercises using two, real, large-scale borehole data
sets involved in the mining of gold and other minerals.  We pit classic
kriging against the modern GPs in several variations and conclude that the
latter can more economical (fewer human and compute resources), more accurate
and offer better uncertainty quantification. We go on to show how the fully 
generative modeling appraratus provided by GPs can gracefully accommodate 
left-censoring of small measurements, as commonly occurs in mining data and 
other borehole assays.
\end{abstract}



\section{Introduction}
\label{sec:introduction}


The modern literature on spatial nonparametric regression (e.g., ``kriging'')
traces its origins to the mining analytics of Danie Krige and Henri de Wijs
and the subsequent work of Matheron \citep{Matheron1971}. Similar ideas were
developed independently around the same time to aid the early analysis of
computer simulation experiments, like those conducted in the study of nuclear
weapons and energy, however (unclassified) publications did not appear until
later (e.g., \cite{Sacks1989}). The spatial statistics community was
responsible for much of the subsequent advances in methodology (e.g.,
\cite{Cressie1993}), and software for kriging in use commercially (e.g., {\tt
LeapFrog}, {\tt Vulcan}, {\tt Surfer}, etc.) and academically (e.g., {\tt
GSLIB}) in mining today.

More recently, researchers in geospatial statistics, surrogate modeling of
computer experiments, and more broadly in statistical and machine learning
communities, have pushed the boundaries of fidelity and computational
tractability as modeling ambition and scale of data collection continue to
grow, e.g., \cite{Gramacy2020}. These disparate literatures have converged
around the nomenclature of Gaussian process (GP) regression as a generative
framework for the kinds of data and procedures involved in kriging, but with a
more cohesive and flexible approach to inference, approximation and automation
based upon the likelihood, which is the foundation to modern statistcal
learning.

Common geoscience applications for spatial smoothing and interpolation include
ore-grade estimation and reservoir characterization/simulation; however,
software tools utilized for such applications lag the state-of-the-art (as
outlined above) by a decade or more. For example, obtaining fits requires
expert human interaction with the software library and intuition in order to
entertain alternatives of spanning anisotropies, neighborhood sizes of
ordinary kriging in the face of large training data sets, and appropriate
semivariogram forms modeling the decay of spatial correlation. Recent advances
from the statistics and data analytics communities automate many of these
time-consuming tasks, while offering substantial improvements in computational
efficiency includeing the use of contemporary computing architectures such as
multi-core workstations and clusters.


The main goal of this paper is to provide a review which focuses on the
similararties and differences in these two frameworks.  Ultimately we wish to
advocate for the more modern, GP perspective via open-source libraries such as
for {\sf R} such as {\tt GPvecchia} \citep{Katzfuss2021} and {\tt laGP}
\citep{Gramacy2016}.  These are just two of many which offer modern take on
ordnary kriging, embodying advances in engineering and statistical learning:
likelihood-based criteria offloaded to robust optimization libraries;
human out-of-the-loop inference.  Not only are they easy to use, but they are
also hard to misuse.

\begin{figure}[ht!]
	\centering
	\includegraphics[height=7cm,trim=10 30 10 50]{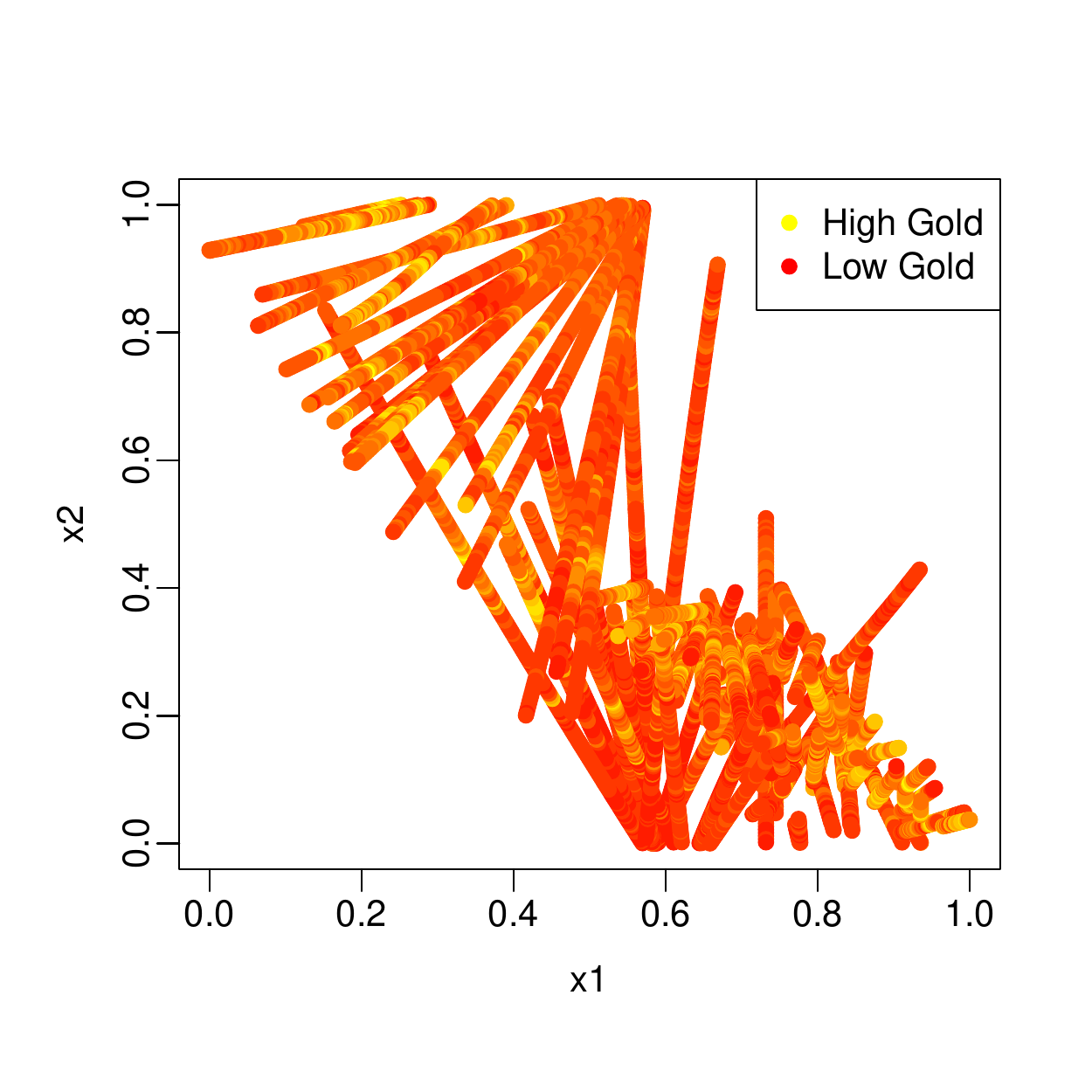}
	\caption{An example of borehole data as a 2d projection.}
	\label{fig:boreholes}
\end{figure}

This narrative is backed up by empirical comparison.  We consider two real
data borehole-based mining examples with data records on gold and other
minerals, over spatial and depth coordinates, sized in the hundreds of
thousands. Figure \ref{fig:boreholes} shows a 2d projection of a subset of the
data described in more detail in Section \ref{sec:results}; note the long
stretches of observations all in a line. These data exhibit many typical yet
challenging features such as abrupt changes in dynamics, left-censoring of
small values due to the sensitivity of the measurement instrument, and large
measurement gaps in space.  Using these data, we devise a
cross-validation-based out-of-sample exercise which is careful to respect the
borehole nature of data collection.  The outcome of that exercise is evidence
that modern GP-based methods are both more accurate, more hands-off, more
economical (in terms of computing resources), and offer better uncertainty
quantification than their kriging-based analogues.  They also enable
extensions which would be difficult to entertain without a fully probabilistic
generative framework.  As a showcase, we entertain a multiple imputation
scheme to handle left censoring that involves only a few lines of code around
library-based GP fitting and prediction subroutines.

The rest of the paper is outlined as follows. Section \ref{sec:review}
contains a review of GP regression and kriging. Building on those, Section
\ref{sec:methods} contains the main, large-scale GP regression and kriging
methods we compare using two real borehole ore data sets. Section
\ref{sec:results} has cross validation results comparing those methods on both
time and accuracy, including extensions to faciliate (without discarding) a
large degree of left-censoring in one of the two data sets. Finally, Section
\ref{sec:conclusion} concludes with a discussion and ideas for future work.

\section{GPs versus kriging} 
\label{sec:review}

We begin by introducing Gaussian process (GP) regression with an eye toward
connecting to kriging.  At some level, they are the same thing.  The biggest
differences lie in vocabulary and inference for unknown quantities, which is
coupled with the degree of automation/human intervention.

\subsection{Gaussian Process Regression}
\label{sec:GPs}
Suppose we wish to model a function $f: \mathbb{R}^d \rightarrow \mathbb{R}$
with a limited number of noisy evaluations $y_i = f(x_i) + \epsilon_i$, for
$i=1,\dots,N$.  Let $X_N$ be an $N \times d$ matrix formed with
$d$-dimensional $x_i^\top$ in each of its rows.  Similarly combine scalar
outputs $y_i$ into an $N$-vector $Y_N$. Throughout this paper, we privilege an
input--output $(x,y)$ notational scheme in keeping with the vast majority
of the statistical learning literature on regression, nonparametric
and nonlinear or otherwise.  In many geospatial contexts
\cite[e.g.,][]{Banerjee2017}, where $f$ might be an environmental or
geological process, it is common to use $s_i$ for (spatial) input sites and $z_i =
Z(s_i)$, among many alternatives, for the response.  We think this unproductively
biases thinking towards $d=2$-dimensional point-referenced (latitude and
longitude) data, whereas these methods can be applied much more widely than 
that. Machine learning \citep[e.g.,][]{Rasmussen2006} and computer surrogate 
modeling \citep[e.g.,][]{Gramacy2020} applications are typically in higher 
input dimension, and one of our goals in this paper is to introduce this way of 
thinking into the mining literature. 


A common nonparametric model for such data is a Gaussian process (GP), which
assumes that outputs $Y_N$ follow a multivariate normal (MVN) distribution.
Inputs $X_N$ are primarily involved in the specification of the MVN covariance
$\Sigma_N \equiv \Sigma(X_N, X_N)$ with a form for $\Sigma(\cdot, \cdot)$ that
inverts Euclidean distances between its arguments.  For example,
\begin{equation}
	Y_N \sim \mathcal{N}_N(0, \Sigma_N),  \quad \mbox{where} \quad \Sigma_N^{ij}
	\quad \mbox{follows} \quad S\left(\frac{1}{\mathrm{Dist}(x_i, x_j)}\right) \quad 
	\mbox{for some decreasing } S.  
	\label{eq:gpprior}
\end{equation}
In Eq.~(\ref{eq:gpprior}) we are being deliberately imprecise about the form
of $\Sigma_N$, a topic we shall detail shortly in Section \ref{sec:modeling}.
For now, simply suppose correlation in outputs decays as a function of
distance in inputs: $\mathbb{C}\mathrm{orr}(y_i, y_j) <
\mathbb{C}\mathrm{orr}(y_i, y_k)$ if $x_i$ is ``closer'' to $x_k$ than it is to 
$x_j$. We are also using a zero mean specification, so that all of the modeling
``action'' is in the covariance.  Extensions abound.

Although a Bayesian interpretation is not essential in characterizing GP 
regression, Eq.~(\ref{eq:gpprior}) can be said to specify a prior over (noisy
evaluations of) functions like $f$, abstracting as $Y_N \sim \mathrm{GP}$.
Choices for the mean (0) and variance ($\Sigma$) determine the modeling
properties of $f$ like its smoothness and wiggliness.  We shall largely leave
those properties to our references, except as relevant to particular choices
for $\Sigma(\cdot, \cdot)$, again in Section \ref{sec:modeling}.  Then, if
$N'$ new locations $\mathcal{X}$ come along where we do not yet have
observations, $Y(\mathcal{X})$, we can summarize our understanding for those in
light of the (training) data we do have -- a predictive distribution --
through the lens of posterior conditioning: $Y(\mathcal{X}) \mid Y_N$.  First,
extend the GP prior to cover $Y(\mathcal{X})$ jointly with $Y_N$:
\[
\begin{bmatrix} Y_N \\ Y(\mathcal{X}) \end{bmatrix} \sim
\mathcal{N}_{N + N'} \left(
\begin{bmatrix}
0 \\ 0 
\end{bmatrix}, 
\begin{bmatrix}
\Sigma_N & \Sigma(X_N, \mathcal{X}) \\
\Sigma(\mathcal{X}, X_N) & \Sigma(\mathcal{X}, \mathcal{X})
\end{bmatrix}
\right).
\]
In so doing, we may leverage that $Y(\mathcal{X})$ values are more highly
correlated with $Y_N$ values whose $X_N$ entries are close to $\mathcal{X}$
ones by applying standard MVN conditioning rules, such as those found in 
\cite{Kalpic2011},
$Y(\mathcal{X}) \mid Y_N \sim \mathcal{N}_{N'} (\mu_N(\mathcal{X}),
\Sigma_N(\mathcal{X}))$ where
\begin{align}
	\label{eq:pred}
	\mu_N(\mathcal{X}) &= \Sigma(\mathcal{X}, X_N)\Sigma_N^{-1}Y_N \\
	\Sigma_N(\mathcal{X}) &= \Sigma(\mathcal{X}, \mathcal{X}) - 
	\Sigma(\mathcal{X}, X_N)\Sigma_N^{-1}\Sigma(X_N, \mathcal{X}). \nonumber
\end{align}
Note that $\Sigma_N(\mathcal{X})$ and $\Sigma(\mathcal{X}, \mathcal{X})$ are
$N' \times N'$ matrices; the $N$ subscript serves as a reminder of
conditioning on $Y_N$.  Observe that $\mu_N(\mathcal{X})$ is a (high
dimensional) linear projection of those $Y_N$ values, where the ``weights''
involved are inversely proportional to the distance between their $X_N$ values
and those of $\mathcal{X}$.  Such conditioning identities apply for any MVN,
based on a GP prior or otherwise.  The special thing about the regression
context is the (inverse) distance-based dynamics manifest as $\mathcal{O}(N)$
weights in each row of $\Sigma(\mathcal{X}, X_N)$, and $\mathcal{O}(N^2)$ in
$\Sigma_N$, involved in the projection, rather than the usual $\mathcal{O}(d)$
or $\mathcal{O}(d^2)$ weights in, say, an ordinary linear regression.  That
higher-dimensional linear projection, $\mu_N(\mathcal{X})$, has properties that
transcend the Bayesian interpretation.  For example, under certain conditions
it is a best linear unbiased predictor (BLUP).  Although much of modern
statistical learning now understands Eq.~(\ref{eq:pred})
in a wider, primarily Bayesian GP context, they are identical to the so-called
{\em kriging equations} \citep{Matheron1963}, which have been instrumental in
geospatial/mining analysis for more than half a century.

To illustrate Eq.~(\ref{eq:pred}) consider $f(x) = 2 +
2\sin(4\pi x)$, observed at $N=20$ $x_i$-values uniform in
$[0,1]$ as $y_i = f(x_i) +
\varepsilon_i$, where $\varepsilon_i \stackrel{\mathrm{iid}}{\sim}
\mathcal{N}(0, 0.1)$. While GP regression is usually applied in higher 
dimension, such as 2d and beyond, the 1d setting is convenient for
visualization.  These twenty $(x_i, y_i)$ pairs comprise of our ``training data''
$(Y_N, X_N)$.  Now, suppose we had a dense testing grid of $N' = 1000$ predictive
locations $\mathcal{X}$ covering $[0,1]$.  Applying Eq.~(\ref{eq:pred}) would
provide us with an $N'$ vector of predictive means $\mu_N(\mathcal{X})$ and an
$N' \times N'$ matrix of predictive covariances $\Sigma_N(\mathcal{X})$
summarizing our regression of $y$ onto $x$.  Variances 
$\sigma^2_N(\mathcal{X})$ along the diagonal of $\Sigma_N(\mathcal{X})$ could 
be used to build error-bars describing a predictive interval (PI) as roughly 
$\mu_N \pm 2\sigma_N$ for 95\% coverage.

\begin{figure}[ht!]
	\centering
	\includegraphics[height=7cm,trim=10 30 10 50]{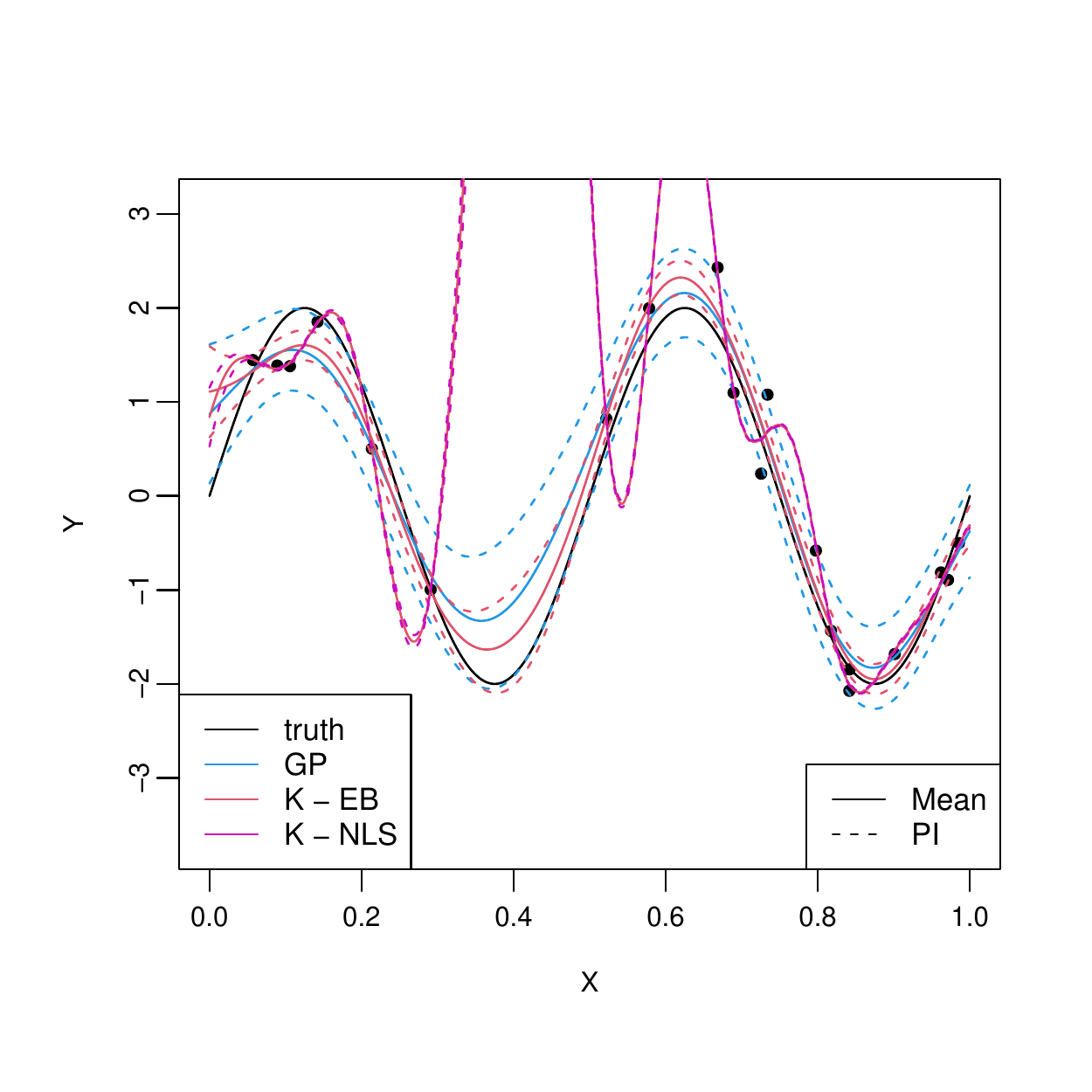}
	\caption{A 1d example function in black with dots being the observed 
	locations. The GP prediction is in blue with dashed lines being the 90\% 
	predictive interval (PI). Kriging predictions are shown in red and magenta 
	with EB and NLS referring to different variography techniques.}
	\label{fig:GP}
\end{figure}

These quantities are shown for one example of such data in Figure
\ref{fig:GP}. Noisy data evaluations (solid dots) dance around the true
unknown function $f$ (black line); our prediction(s) $\mu_N$ and PIs in
(blue/red lines, dashed respectively) in two variations (labeled ``GP'' and
``Kriging'') accurately distill the essence of the input-output relationship.
All of this is modulo a fortuitous choice for $\Sigma(\cdot,
\cdot)$ which we have yet to detail.  Its specification, and method of
inference or unknown quantities, comprises the wedge between modern GPs (red in
the figure) and traditional kriging (blue).

\subsection{Modeling}
\label{sec:modeling}

The discussion above hinges on a choice of $\Sigma(\cdot, \cdot)$, or $S$ in
Eq.~(\ref{eq:gpprior}).  $S$ was merely used as a notational device to delay
discussion until this moment; we shall not use $S$ going forward. Yet, that
Eq.~(\ref{eq:gpprior}) formulation is attractive because it abstracts all
modeling details down to this ``one'' choice. ``One'' is in quotes because
$\mathcal{O}(N^2)$ quantities, one for each pair of $N$ data elements $(x_i,
y_i)$, are actually constrained 
by the covariance structure.  This vast number of potentially tunable
quantities, more even than $N$, is why one refers to GPs as nonparametric. But
of course, it's neither practical nor valid to allow oneself such unbridled
freedoms. For example, we must choose $\Sigma(\cdot, \cdot)$ so that
$\Sigma_N$ is finite and positive definite for use as an MVN covariance.

An inverse-distance-based covariance is conventional as an intuitive spatial
modeling device.  However, this is not a requirement and may not be ideal in
all situations, e.g., when modeling periodic effects.  We may wish to allow
flexibility in how distances are measured, in what coordinates and with what
decay in inversion, and to control how such choices relate signal to noise.
Such considerations lead to frameworks for choices of $\Sigma(\cdot, \cdot)$
whose tunable quantities, sometimes called {\em hyperparameters} to
acknowledge a nonparametric modeling apparatus, can be learned from data.

For example, if the range of the responses $Y_N$ is unknown {\em a priori} we
might wish to design $\Sigma(\cdot, \cdot)$ to include a scale hyperparameter,
say $\tau^2$.  If $Y_N$ is noisy and/or contains measurement error, we may
wish to encode it as part signal, and part noise.
Sometimes this is governed by a so-called {\em nugget} hyperparameter, which
we shall denote as $g$.  We caution that the role of GP nugget is inspired by,
but is subtly different from, a parameter of the same name in the
geostatistics/kriging literature.  More in Section \ref{sec:kriging}.  We
may wish to control the smoothness and rate of decay of correlation of the
signal in terms of (inverse) distance, and thereby the smoothness and other
properties of the underlying response surface.  This may be accomplished
through selection of a so-called {\em kernel} function $k_\theta(x_i, x_j) :
\mathbb{R}^d \rightarrow [0,1]$ whose hyperparameter $\theta$ can be used to
describe the rate of radial decay from $k_\theta(x_i, x_j) = 1$ for $x_i =
x_j$ down to zero as $x_j$ moves away from $x_i$ in an isotropic modeling
context. Kernels $k$ may also re-scale and/or rotate the space for anisotropic
effects. 

One way to put these elements together is
\begin{equation}
\label{eq:sigma}
\Sigma(x_i, x_j) = \tau^2(k_\theta(x_i, x_j) + g \delta_{ij})
\quad \mbox{ so that } \quad \Sigma_N = \tau^2(K_N + g\mathbb{I}_N).
\end{equation}
Above, $\delta_{ij}$ is the Kronecker delta function returning 1 when the
index $i = j$, i.e., when the same training data element appears in both
arguments, and zero otherwise, and $K_N \equiv k_\theta(X_N, X_N)$ applying
$k$ elementwise as $K^{ij}_N = k_\theta(x_i, x_j)$.  Observe that the
diagonal of $\Sigma_N$ is $\tau^2 (1 + g)$ and all off-diagonal entries are less
than or equal to $\tau^2$, and strictly less than for all $x_i \ne x_j$.  This
discontinuity between diagonal and off-diagonal, as long as $g > 0$, leads to
smoothing of the predictive surface when following Eq.~(\ref{eq:pred}).
Otherwise, when $g=0$ the surface interpolates.  Again, this is a little
different than the typical geostatistics formulation as explained later in
Section \ref{sec:kriging}.

Choices for distance-based kernels $k$ preserving positive definiteness and
targeting certain other properties abound.  See, e.g.,
\citet{abrahamsen1997review} or \citet{wendland2004scattered}.  The two that
are most widely used are the power exponential and the Mat\'ern.  These are
provided below in an isotropic setting, i.e., with radial decay as a function
of distance.
\begin{align}
	\label{eq:kernels}
	\begin{split}
		k_\theta^{P}(x_i, x_j) &= \exp\left\{-\frac{||x_i - 
		x_j||^p}{\theta}\right\}\\
		k_\theta^{M}(x_i, x_j) &= \frac{2^{1 - \nu}}{\Gamma(\nu)}\left(||x_i - x_j||
		\sqrt{\frac{2\nu}{\theta}}\right)^{\nu}\mathcal{K}_{\nu}\left(||x_i -
		x_j||\sqrt{\frac{2\nu}{\theta}}\right)
	\end{split}
\end{align}
In both cases above the hyperparameter $\theta$ appears in the denominator,
scaling (squared or square root) Euclidean distances between $x_i$ and $x_j$,
and is thus sometimes called the {\em characteristic lengthscale}.  Both have
additional hyperparmeters, $p$ and $\nu$, respectively, which must be positive
and which our notation does not include in $\theta$ because they are usually
specified, rather than inferred with more discussion provided in Section
\ref{sec:conclusion}. These control the of the resulting response surface.
When $p=2$ the power exponential produces infinitely smooth (mean-square
differentiable) realizations, and sometimes this special case is called the
Gaussian kernel,\footnote{{\em Gaussian} here refers to the expression
resembling the density of a Gaussian distribution; it has nothing to do
with making a Gaussian assumption, or its use in GP. Such kernels are used in
a variety of other contexts.} which we denote as $k_\theta^G$. When $p \ne 2$,
the surface is nowhere differentiable. While sometimes such pathological
non-smoothness is a reasonable assumption -- and in spite of this the
predictive surfaces (\ref{eq:pred}) often look smooth
-- there are better mechanisms for relaxing
unreasonable (infinite) smoothness.

Many of the choices in our references above offer higher fidelity control over
smoothness (beyond none and infinite).  Of those, the Mat\'ern has percolated
into the canonical position thanks largely to persuasive technical arguments
from \citet{stein1999interpolation}.  The parameter $\nu$ controls this
aspect, with higher values leading to greater smoothness, yielding surfaces
which are $\lceil \nu \rceil - 1$ mean-square differentiable. Ultimately when
$\nu \rightarrow \infty$ the Gaussian kernel is recovered as a special case. 
However, the modified Bessel function $\mathcal{K}_\nu$ can be 
difficult to
work with computationally. Specific settings with $\nu \in \left\{\frac{3}{2},
\frac{5}{2}\right\}$ have algebraic closed forms (no Bessel functions) which
yield degree one and two differentiability, with the latter being most often
applied in practice because most interesting dynamics result from at least
twice (but not infinitely) differentiable processes.

Both the (isotropic) power exponential/Gaussian and Mat\'ern are {\em
stationary} kernels because they are defined only in terms of displacement
$||x_i - x_j||$, so the resulting response surface would have identical dynamics
throughout the entire input space.  Nonstationary modeling is also possible,
but is a lot more difficult in general.  See, e.g.,~\citet{sauer2021active}, and
further discussion in Section \ref{sec:transductive}.  However, one can
still capture nontrivial dynamics with stationary kernels, for example by
deploying several of them simultaneously: sums, products, convolutions (and
more) of valid kernels for GP regression (i.e., are positive definite) are
also valid.  For details, see e.g., \citet[][, Section 4.2.4]{Rasmussen2006} or
\citet[][, Section 5.3.3]{Gramacy2020}.  One of the most common applications
of this result is to extend to axis-aligned anisotropy by taking a product of
kernels applied univariately in each coordinate direction: $k_\theta(x_i, x_j)
= \prod_{k=1}^d k_{\theta_k}(x_{ik}, x_{jk})$, abusing the notation somewhat.
This can be done with any kernel. Notice here we are introducing a
$d$-dimensional lengthscale parameter $\theta = (\theta_1,\dots,\theta_d)$,
controlling the rate of spatial correlation differentially in each coordinate
direction.  For the Gaussian kernel, the result is identical to
\begin{equation}
	\label{eq:anisotropy}
	k_\theta^G(x_i, x_j) = \exp\left\{-\sum_{k = 1}^d\frac{(x_{ik} - 
		x_{jk})^2}{\theta_k}\right\},
\end{equation}
which is sometimes called the {\em separable} Gaussian kernel or the ARD
Gaussian kernel.  ARD stands for {\em automatic relevance determination}
\citep{Liu2020}, borrowing terminology from early neural networks literature.
The idea is that the data can inform on longer lengthscales (less relevant) or
shorter ones (more relevant) for each input variable separately.
ARD/separable  Mat\'ern kernels are also common, but their expression(s) are
less tidy so we do not include it here.  For more discussion, consult
\citet{Rasmussen2006} or \citet{Gramacy2020}.

There is a one-to-one relationship between vectorized lengthscale in the ARD
kernel formulation and with re-scaling inputs $X_N$, say as a pre-processing
step.  Rather than scaling each input differently, one can extend this idea
to rotations and projections to accommodate less rigid anisotropy either as
preprocessing or as a hyperparameterized kernel formulation.  For an example
of a pre-processing approach see \citet{wycoff2021sensitivity} and references
therein; for learning rotations and projects {\em within} a kernel see
\citet{gramacy2012gaussian}.  We find that such high-powered approaches are
overkill for most applications, including the mining ones discussed later.

\subsection{Inference}
\label{sec:inference}

The models above have tunable quantities, or hyperparameters, that
could be set by hand but would ideally be learned from data.  We restrict our
focus to those which we introduced for $\Sigma(\cdot, \cdot)$, particularly
$\phi \equiv (\tau^2, g, \theta)$ with the latter usually vectorized in an ARD
setting, but there could potentially be additional quantities which
must be estimated from data.  There are many criteria and algorithms across
several literatures devoted to such ``fitting'' enterprises
\citep[e.g.,][]{Diggle2007}. Yet there is a remarkable confluence in modern
statistical learning practice when it comes to the near universality of
likelihood-based methods when distributional assumptions are being made (like
the MVN in Eq.~(\ref{eq:gpprior})).  The reason is that no additional criteria
need be introduced to commence with learning. One may choose to impose
additional assumptions, like priors on aspects of $\phi$ for a Bayesian
approach \citep[][which is still likelihood-based]{Banerjee2017}, or not --
simply maximize the likelihood. This is the approach we present here because
it is tidy and fast.

Eq.~(\ref{eq:gpprior}) depicts how observations/outputs (like $Y_N$), are
distributed in relation to inputs (like $X_N$) and parameters or
other structure (like $\Sigma(\cdot,\cdot)$ via hyperhaprameters $\phi$ and
kernels $k$).  The {\em likelihood} simply re-frames the density of that
distribution, which assigns positive real values to $Y_N$ as a function of
parameters (or hyperparameters $\phi$, say), the other way around: providing
positive reals for $\phi$ given $Y_N$.  Once in that context, it makes sense
to seek out the parameterization that makes the observed $Y_N$ most likely,
i.e., that maximizes the likelihood.  There are two benefits to this approach.
One is that it reduces a statistical inference question to an optimization
one without introducing auxiliary criteria.  The other is that the solution
to this optimization, the so-called maximum likelihood estimator (MLE), has
special properties that can be used to quantify uncertainty. For a review of
likelihood-based inference, see \citet{CaseBerg:01}.

By inspecting the MVN density \citep{Kalpic2011}, using a mean of zero and 
covariance $\Sigma_N$, one may obtain the following expressions for the 
likelihood and its logarithm.
\begin{align}
		L(\phi; Y_N) &= 
		(2\pi)^{-\frac{N}{2}}|\Sigma_N|^{-\frac{1}{2}}\exp\left\{ 
		-\frac{1}{2}Y_N^\top \Sigma_N^{-1}Y_N\right\}\label{eq:gpll}\\
		\ell(\phi; Y_N) = \log(L(\phi; Y_N)) &= -\frac{N}{2}\log 2\pi  - 
		\frac{1}{2}\log |\Sigma_N| - \frac{1}{2}Y_N^\top \Sigma_N^{-1}Y_N
		\nonumber
\end{align}
Recall from Eq.~(\ref{eq:sigma}) that the parameters $\phi$ are embedded in
$\Sigma_N = \tau^2(K_N + g\mathbb{I}_N)$ via $K_N$ which is built through
pairwise application of $k_\theta(\cdot, \cdot)$. Technically, this is a {\em
marginal likelihood} since it tacitly defines both prior (which is integrated
out) and observational variability.  But this distinction is not important for
our discussion.  The curious reader is referred to Section 5.3.2 of
\citet{Gramacy2020}.  The log likelihood helps with maximization, easing
differentiation but preserving critical points.

Consider $\tau^2$ first.  Observe that 
\[
\ell(\phi; Y_N) = c_1 - \frac{N}{2} \log \tau^2 + \frac{N}{2\tau^2} Y_N^\top \Sigma_N^{-1}Y_N
\]
where $c_1 = - \frac{N}{2} \log 2\pi - \frac{1}{2} \log |K_N + g\mathbb{I}_N|$ is
constant with respect to $\tau^2$.  Differentiating with respect to $\tau^2$,
setting to zero and solving, is straightforward.
\begin{equation}
	\label{tau_deriv}
		0 \stackrel{\mathrm{set}}{=} \frac{d\ell(\phi; Y_N)}{d\tau^2} = -\frac{N}{2\tau^2} + 
		\frac{1}{2(\tau^2)^2}Y_N^\top (K_N + g\mathbb{I}_N)^{-1} Y_N^\top \quad \rightarrow \quad
		\hat{\tau}^2 = \frac{Y_N^\top (K_N + g\mathbb{I}_N)^{-1} Y_N}{N}
\end{equation}
If we then plug $\hat{\tau}^2$ back into $\ell(\phi; Y_N)$ we get what is known
as a concentrated, or profile log likelihood:
\[
\ell(\theta, g; Y_N) = \ell(\phi; Y_N) \big{|}_{\tau^2 = \hat{\tau}^2} = 
c_2 - \frac{N}{2} \log Y_N (K_N + g\mathbb{I}_N)^{-1} - \frac{1}{2} \log | K_N + g\mathbb{I}_N|
\]
where $c_2$ is not a function of $\theta$ or $g$.  Differentiation here is
more challenging because the parameters are buried within matrix inverses and
determinants.  But it is still doable.  See, e.g., Eq.~(5.9) in 
\citet{Gramacy2020}.  We have
\begin{align}
		\frac{d\ell(\theta, g; Y_N)}{dg} &= \frac{N}{2}\frac{Y_N^\top((K_N + g 
		\mathbb{I}_N)^{-1})^2Y_N}{Y_N^\top (K_N + g\mathbb{I}_N)^{-1}Y_N} - 
		\frac{1}{2}\mathrm{tr}((K_N + g \mathbb{I}_N)^{-1}) 
		\label{eq:other_deriv}\\
		\frac{d\ell(\theta, g; Y_N)}{d\theta} &= \frac{N}{2}\frac{Y_N^{\top}K_N^{-1} K_N' (K_N + g \mathbb{I}_N)^{-1}Y_N}{Y_N^{\top} (K_N + g \mathbb{I}_N)^{-1}Y_N} - \frac{1}{2}\mathrm{tr}((K_N + g \mathbb{I}_N)^{-1}K_N'), \nonumber
\end{align}
where $K_N' = \frac{dK_N}{d\theta}$. The former is actually a special case of
the latter, taking instead $\theta$ as $g$, since $\frac{d(K_N +
g\mathbb{I}_N)}{dg} = \mathbb{I}_N$. When $\theta$ are lengthscales in 
$k_\theta$, $K_N'$ is formed of $(K_N')^{ij} =
d k_\theta(x_i, x_j)/d\theta$, i.e., with component-wise derivatives
of the kernel, say Gaussian or Mat\'ern, with respect to $\theta$.  When
$\theta$ is vectorized, we may instead calculate $\partial \ell(\theta, g; 
Y_N) / \partial \theta_k$, forming a gradient for $k=1,\dots,d$.

The next step is to set these derivatives to zero, either simultaneously or
separately, and solve.  However, no algebraic solution is known.
Library-based numerical optimization, such as BFGS \citep{Byrd1995} provided
in {\sf R}'s {\tt optim} function, can be used to find MLEs $\hat{g}$ and
$\hat{\theta}$. Convergence is usually both robust and fast, even in high
dimension $d$, when gradients (\ref{eq:other_deriv}) are provided since the
underlying $\ell(\theta, g; Y_N)$ surface is almost always convex.
\begin{figure}
	\includegraphics[width=7cm,trim=10 30 10 20]{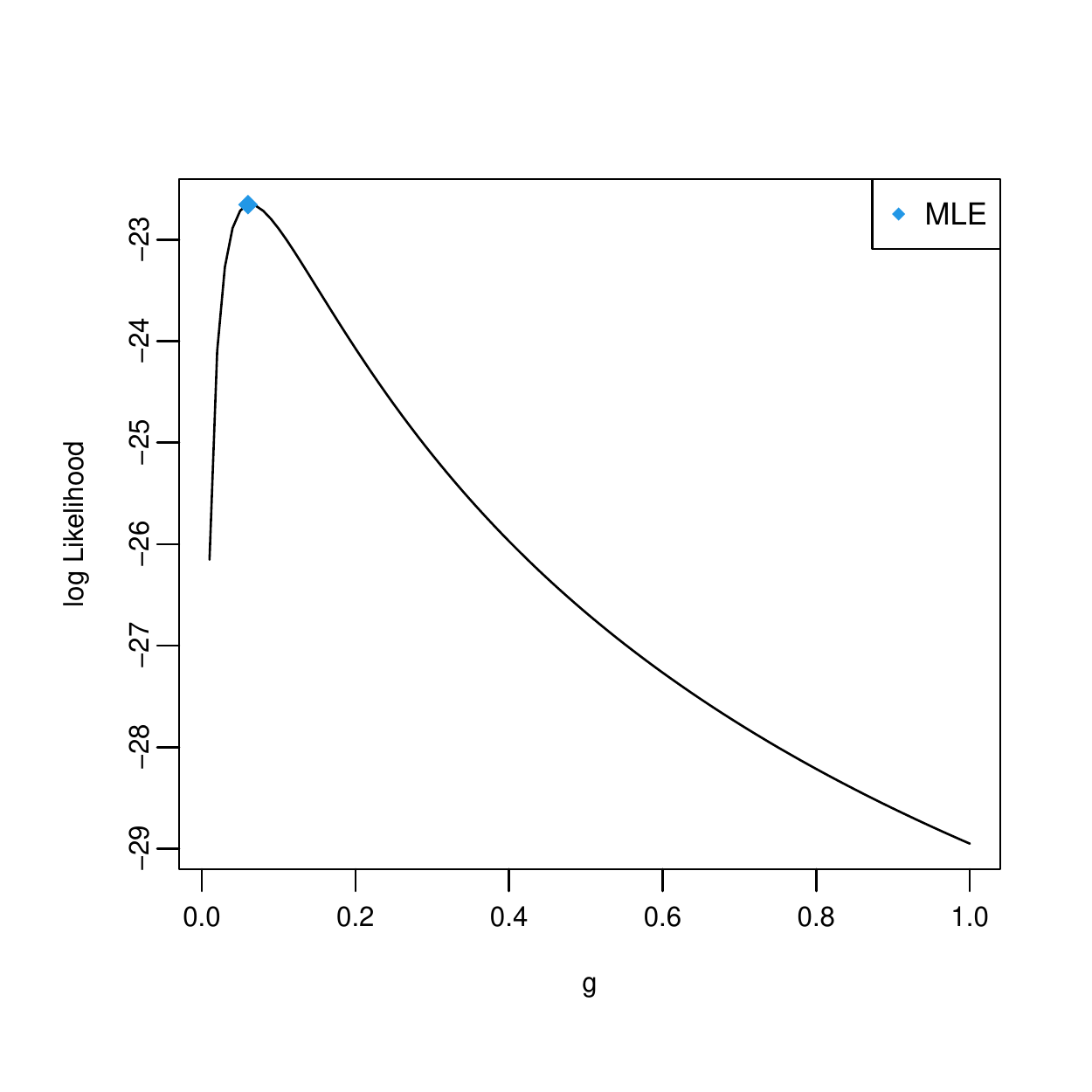}
	\includegraphics[width=7cm,trim=10 30 10 20]{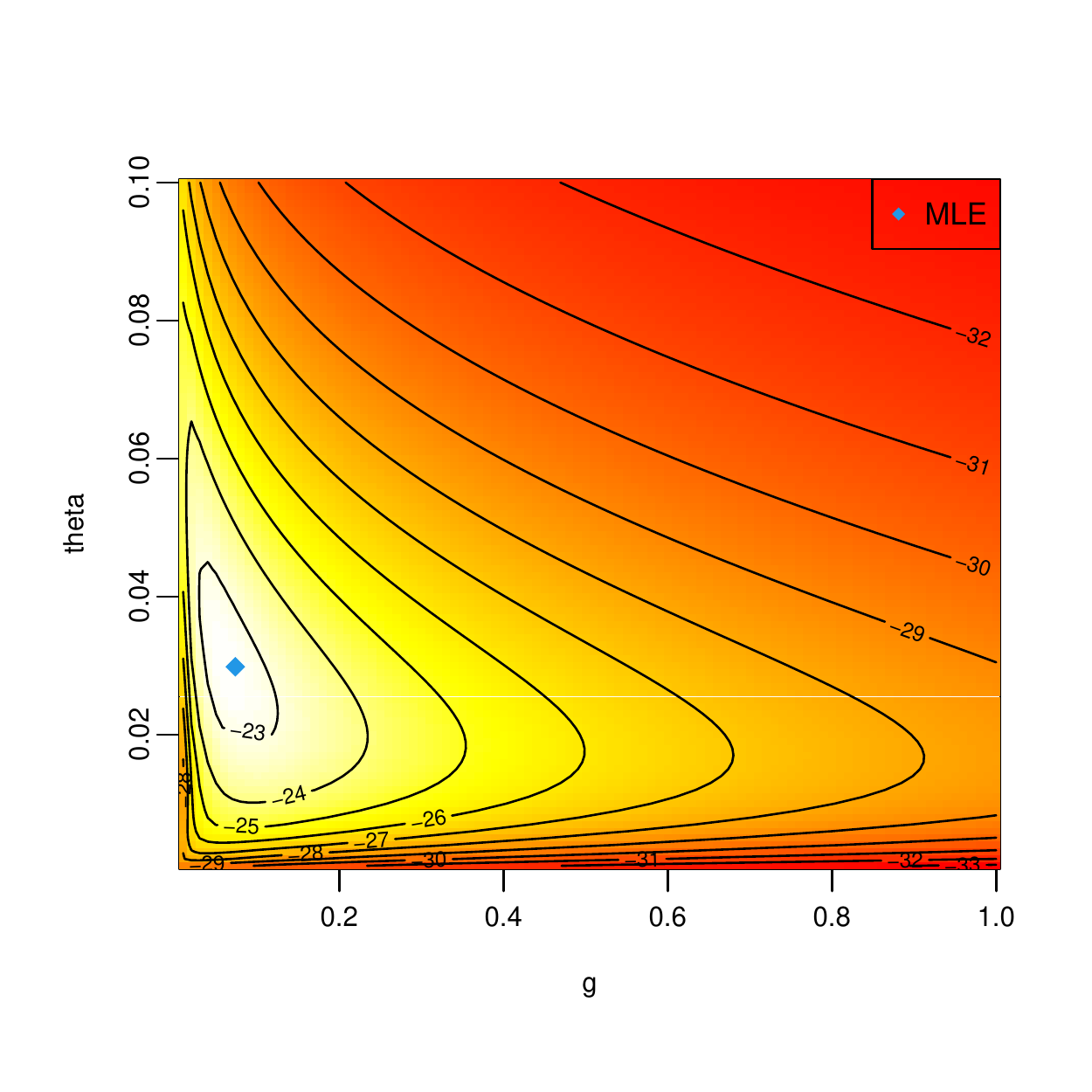}
	\caption{Log likelihood surfaces for the function in Figure \ref{fig:GP}.
	Left: Surface for $g$, holding $\tau^2$ and $\theta$ constant at their MLE
	values. Right: Surface for $g$ and $\theta$, holding $\tau^2$ at its MLE
	value.}
	\label{fig:loglik}
\end{figure}
To illustrate, Figure \ref{fig:loglik} shows  $\ell(\theta, g; Y_N)$ using
data associated with Figure \ref{fig:GP}. The left panel varies $g$
conditional on a setting of $\theta$, whereas the right panel varies both.  In
both cases, $\hat{\tau}^2$ has been concentrated out.  Note that raw
(un-logged) likelihood surfaces are even more peaked than these figures
present. Regardless, it is easy to eyeball the MLE.  {\sf R}'s {\tt optim} is
able to find $\hat{g}$ and $\hat{\theta}$, shown as blue diamonds in both
panels in seventy-nine iterations.  Looking back at Figure \ref{fig:GP}, the
blue curves correspond to predictive surfaces uses these MLE hyperparamters.
The red and magenta curves are detailed in the next subsection.


\subsection{Kriging and Variography}
\label{sec:kriging}

The main difference between classical kriging and the GP presentation above
regards inference for unknown quantities and, in the case of the latter, a
more up-front and highly-leveraged distributional assumption (Gaussian) for
the response.  Both use Eq.~(\ref{eq:pred}) to form predictions and quantify
uncertainty. In geostatistics,
these are known as the ``kriging equations,''
even when other aspects historically associated with kriging are not
faithfully replicated. Classical kriging focuses on lower input dimension
-- particularly $d \in \{2,3\}$ in spatial contexts -- and as such prefers
   isotropic modeling after a suitable transformation of spatial inputs. {\em
   Variography} is used to select the kernel and its hyperparameters, rather
   than the likelihood.  This has advantages and disadvantages. Many of the
   advantages are related to the historically larger training data sets
   encountered in spatial problems, although that gap is narrowing in wider
   statistical learning and computer experiments contexts. More on this in
   Section \ref{sec:methods}, wherein further distinctions arise. The main
   disadvantage is that input pre-processing and variogram inspection are
   inherently hands-on, human driven enterprises, albeit ones enhanced by
   computational tools. Other differences are more superficial, like naming,
   symbol choice and applications of hyperparameters within variography.

The \emph{semivariogram}, or half the {\em variogram}, denoted as $\gamma(h)$,
 is the variance of two output $y$-values that are distance $h$ apart in the
 input $x$-space: $\gamma(h) = \frac{1}{2}\mathbb{V}\mathrm{ar}(Y(x + h) -
 Y(x))$. Implicit in this definition is an assumption of intrinsic
 stationarity, implying that $\mathbb{E}[Y(x + h) - Y(x)] = 0$, or that the
 covariance between two $y$ values depends not on position but on relative
 distance notated by the displacement $h$ between them. If $h$ is calculated
 using Euclidean distance, intrinsic stationarity implies isotropy. When this
 is a limitation to effective spatial modeling, one may prescale or rotate the
 coordinate system.  This is often based on expert-judgment of the prevailing
 variabilities within the input domain, like the direction of an ore body
 within the geologic topology.  As mentioned earlier, a modern GP approach
 would deploy separable lengthscales (\ref{eq:anisotropy}), or more flexibly
 parameterized rotations and scales that are learned jointly with other
 unknowns.

\begin{figure}
	\begin{minipage}{0.49\textwidth}
		\centering{}
		\includegraphics[height=6.5cm,trim=10 30 10 20]{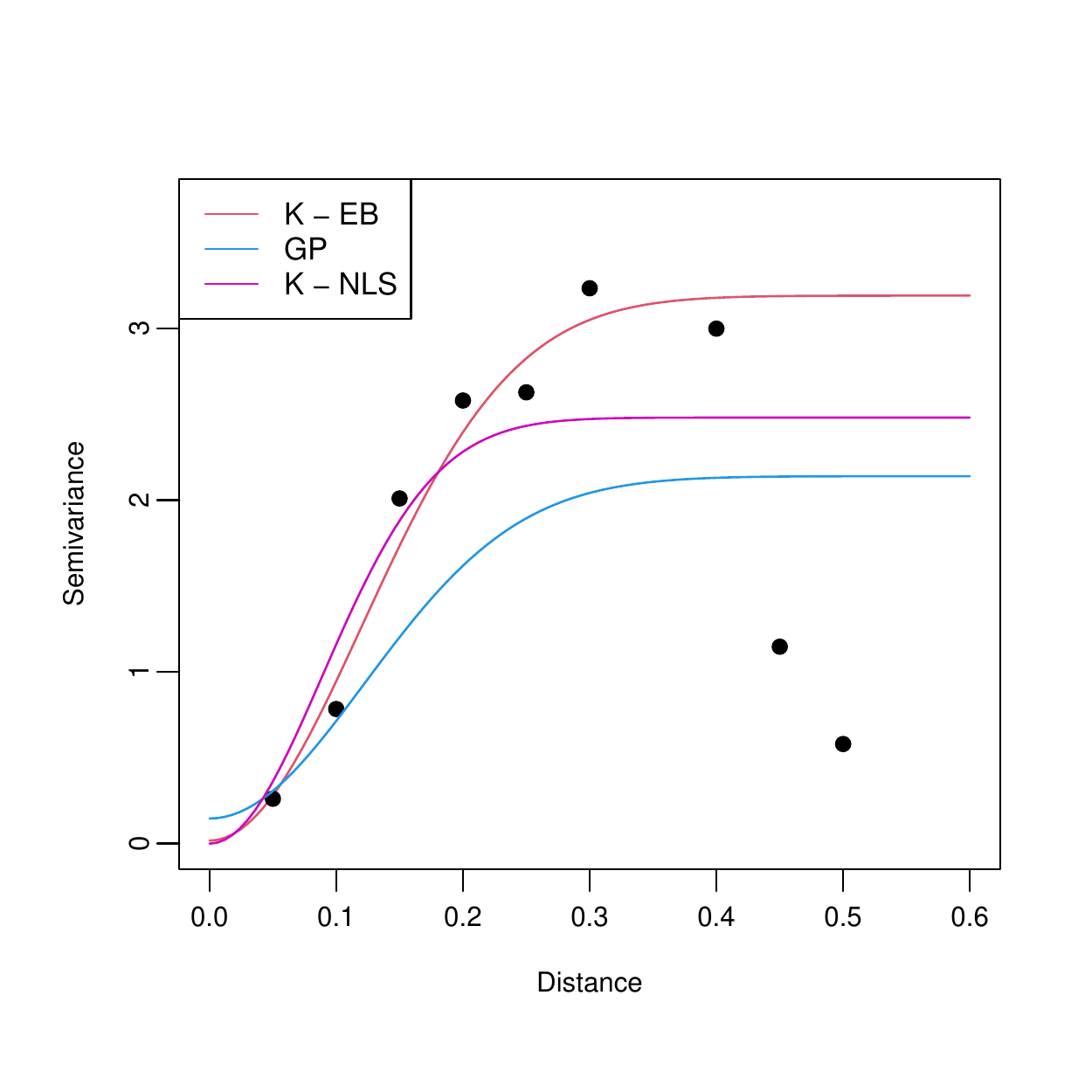}
		\caption{A Gaussian kernel/semivariogram fit to the function in Figure \ref{fig:GP}.  
		``EB'' denotes a semivariogram fit by hand, whereas ``NLS'' uses 
		non-linear least squares;
		``GP'' derives the semivariogram from an  MLE hyperparameterization. 
		 The table compares hyperparameter estimates.}
		\label{fig:Vgram}
	\end{minipage}
	\centering
	\begin{minipage}[t]{0.39\textwidth} \footnotesize
 		\begin{tabular}{l|l|l|l|}
			& GP & K-EB & K-NLS\\
			\hline
			$\tau^2$ or $\sigma^2$ & $1.99$ & $3.17$ & $2.48$\\
			\hline
			$\tau^2g$ or $\tau^2_k$ & $0.145$ & $0.018$ & $0.00010$\\
			\hline
			$\theta$ or $R$ & $0.17^2$ & $0.17$ & $0.13$\\
			\hline
		\end{tabular}
	\end{minipage}
\end{figure}

The semivariogram is a theoretical/population construct that would be hard to
specify {\em a priori} even with expert knowledge, but simple to observe
empirically given data.  One estimate of an {\em empirical semivariogram}
could be obtained by binning the data by distance and calculating sample
covariances within those bins.  Let $N(h_k) = \{(x_i,  x_j): ||x_i - x_j|| \in
I_k\}$ where $I_1 = [0, h_1], I_2 = (h_1, h_2], \dots, I_k = (h_{k - 1}, h_k]$
denote a neighborhood structure striated by bands of distance
$0,h_1,\dots,h_k$.  Then estimate
\begin{equation}
	\hat{\gamma}(h) = \frac{1}{2|N(h)|} \sum_{(x_i, x_j) \in N(h)} (y_i - 
	y_j)^2.
\label{eq:semivar}
\end{equation}
As defined continuously for any $h$, $\hat{\gamma}(h)$ is a step function.
However it is customarily visualized discretely as a scatter plot with $(h_i +
h_{i+1})/2$ as the $x$-axis coordinate.  The left panel of Figure
\ref{fig:Vgram} shows these as dots for the 1d problem  
introduced in Figure \ref{fig:GP} using a bin size ($h_{i + 1} - h_i$) of 0.05.

One can then match these empirical observations of spatial covariance with a
parameterized form for the population semivariogram.  Here, similar constructs
are used to model spatial dependence as the kernels introduced earlier
(\ref{eq:kernels}).  Let $\gamma(0) = 0$ and for $h > 0$, power exponential
and Mat\'ern model semivariograms are often written as
\begin{align}
	\label{eq:vgrams}
	\begin{split}
		\gamma_\theta^{P}(h) &= \tau^2_k + \sigma^2
		\left(1- \exp\left\{-\left(\frac{h}{R}\right)^p\right\}\right)\\
		\gamma_\theta^{M}(h) &= \tau^2_k + \sigma^2\left( 1 - \frac{2^{1 - 
		\nu}}{\Gamma(\nu)}\left(h\sqrt{\frac{2\nu}{R}}\right)^{\nu} 
		\mathcal{K}_{\nu}\left(h\sqrt{\frac{2\nu}{R}}\right)\right).
	\end{split}
\end{align}
Semivariogram parameters are known as {\em nugget} 
($\tau_k^2$),\footnote{A subscript $k$ is not standard; we added it 
to distinguish with the GP scale $\tau^2$.} {\em partial 
sill} ($\sigma^2$), and {\em range} ($R$). 

Taking $\gamma(0) = 0$ is a contentious choice outside of the geospatial
modeling literature.  It implies that there is no intrinsic variance in
measurements.  In part this is because such measurements are inherently
unrepeatable in certain contexts; you cannot dig a borehole in the same place
twice.  But if you could, it stands to reason that you would get different
measurements for such {\em replicates} for all sorts of reasons, e.g., even
without operator error the drill bit might interact with the surface and ore
body differently the second time. Primordial process producing the ore body
are subject to uncertainties that are best characterized as random variables
even if the process is not inherently stochastic. Thus, it is acknowledged that
there will be small-scale variability between {\em nearby} observations that
are best described by noise.  This noise, at all distances $h = \epsilon > 0$,
is what is parameterized by the nugget $\tau^2_k$. The distinction with the GP
nugget $g$, which characterizes the noise as $\tau^2 g$ at $h=0$, is thus
subtle. Choosing $\gamma(0) = 0$ has an impact on the kriging equations
(\ref{eq:pred}), leading to discontinuities at the training data locations,
where an otherwise smooth predictive surface would be pocked with spikes of
``interpolation.'' We do not show these in the red curve by deliberately
omitting $X_N$ values from our predictive grid $\mathcal{X}$ for Figure
\ref{fig:GP} for aesthetic reasons.

Kriging-versus-GP distinctions between the other two kernel parameters are
more superficial. The partial sill $\sigma^2$ controls the maximum covariance 
as $h \rightarrow \infty$. This has a 1:1 correspondence with the $\tau^2$
from earlier. Sometimes the \emph{sill} parameter, $\tau_k^2 +
\sigma^2$, is preferred by geostatisticians instead. The range $R$ controls the 
distance between maximum and minimum covariance, and plays an identical role as
the square root lengthscale: $R = \sqrt{\theta}$. It is not uncommon to
instead specify a decay parameter $\phi = 1/R$, and such inversions are common
in the GP literature as well.

Each setting of these parameters could be used to overlay a curve onto the
left panel of Figure \ref{fig:Vgram}.  For example, using the MLE hyparameters
from our earlier GP analysis yields the curve in blue. Alternatively, one
could automate a search for the ``best fitting'' variogram parameterization
with a generalized/nonlinear (possibly weighted) least-squares (NLS) criterion
\citep{Cressie1985}.  This corresponds to the magenta curve. Observe that
neither of these result in a terrific fit to the semivariogram ``data.'' An
outlying pair of dots near $h=0.5$ drags these variograms down, sacrificing
fit for smaller pairwise distances.  One reason these are outlying may be that
we have many fewer long-distance pairs in the data than short distance ones. A
common remedy would be to downweight or altogether ignore these when fitting
the variogram parameters, focusing  only on the short distance readings.  We
refer to one such fit as the ``eyeball'' (or EB) variogram in the figure,
although in practice NLS may similarly be deployed. This can lead to more
accurate predictions out-of-sample, as we demonstrate momentarily.  However it
has the downside of introducing non-statistical (e.g., NLS) and non-metric
(determination of outlying semivariogram estimates) criteria which diminishes
reproducibility and automation, and incurs the expense of human expert
intervention.  This enterprise is also sensitive to other choices such as bin
size $h_{i+1} - h_i$ and a choice of maximum distance to calculate the
empirical variogram.   We chose $h_{\max} = 0.5$ for Figure \ref{fig:Vgram},
but could have gone out to $h_{\max} = 1$, producing a much ``noisier''
empirical semivariogram.

Figure \ref{fig:Vgram} details hyperparameter estimates for each of the three
techniques. Lengthscale and range settings exhibit high agreement. For
scale/partial sill, NLS and GP MLE values are ``drawn down'' by the noisier
higher distance bins relative to our EB alternative which ignored those
values. The nugget is where things start to substantially diverge: $\tau^2g
\gg \tau^2_k$ means our GP-MLE estimates more noise/less signal than the
kriging alternatives (EB and NLS). Notice that EB and NLS nuggets are on
different orders of magnitude.  The tiny NLS $\tau^2_k$ may be attributed to a
lack of small distance pairs, and consequently the optimizer converged at the
boundary of our search space for that parameter: $10^{-4}$, meaning very high
signal/low noise. Although this seems innocuous when it comes to the 
corresponding semivariograms on the left in the figure, the implications
out-of-sample are severe.  This is foreshadowed in Figure \ref{fig:GP}, and 
evaluated empirically next.

\subsection{Out of Sample Validation}
\label{sec:oosv}

In Figure \ref{fig:GP} the EB kriging fit (solid-red) is visually similar to
the GP-MLE fit (solid-blue) except perhaps near $x=0.4$. This is noteworthy in
light of the disparate parameterization and semivariograms in Figure
\ref{fig:Vgram}, and in particular the human intervention required to ignore
outlying values in favor of short distances. Qualitatively, the red curve may
be more accurate compared to the truth (black), but closer inspection reveals
a more pernicious concern despite apparent higher accuracy: poor uncertainty
quantification.  The red 90\% PI (error-bars) cover only about half of the
training data locations, suggesting that nominal coverage has not been
achieved. In contrast, the blue (GP-MLE) error-bars cover many more of the
data points.

There are many ways to be more precise about out-of-sample prediction
accuracy. Consider a testing set comprised of a grid of $\mathcal{X}$-values
of size $N'$ with true values $y(\mathcal{X})$. Given predictions from a
kriging/GP fit (\ref{eq:pred}), notated generically as $\mu(\mathcal{X})$, the
{\em proper scoring rule} \citep{Gneiting2007} is root mean square error
(RMSE).
\begin{equation}
	\text{RMSE}(y(\mathcal{X}), \mu(\mathcal{X})) = \sqrt{\frac{1}{N'}\sum_{i = 
			1}^{N'} (y_i(x_i) - \mu_i(x_i))^2}
	\label{eq:rmse}
\end{equation}
Lower RMSE is better. The first row of the left panel of Table
\ref{tab:toyscores} shows that the mean predictions for kriging via EB are
indeed more accurate than via MLE.
\begin{table}
	\begin{tabular}{l|l|l|l|}
		1d & GP & K-EB & K-NLS\\
		\hline
		RMSE & 0.14 & 0.084 & 23.11\\
		\hline
		score$_\mathrm{f}$ & 869 & -4161 & -1255140\\ 
		\hline
	\end{tabular}
	\hspace{1cm}

	\caption{Out-of-sample RMSE and score for GP and kriging on the 
	1d toy problem.}
\label{tab:toyscores}
\end{table}

If one also has covariances $\Sigma(\mathcal{X})$, or just variances
$\sigma^2(\mathcal{X}) = \mathrm{diag}(\Sigma(\mathcal{X}))$, associated with
predictions, the proper scoring rule is more complex as accuracy must be
normalized by uncertainty:
\begin{align}
	\text{score}_{\mathrm{f}}(Y(\mathcal{X}), \mu(\mathcal{X}), \Sigma(\mathcal{X})) &= 
	-\log(|\Sigma(\mathcal{X})|) - (Y(\mathcal{X}) - \mu(\mathcal{X}))^\top 
	\Sigma(\mathcal{X})^{-1} (Y(\mathcal{X}) - \mu(\mathcal{X}))
	\label{eq:score}\\
	\text{score}_{\mathrm{p}}(Y(\mathcal{X}), \mu(\mathcal{X}), \sigma^2 (\mathcal{X})) &= 
	-\sum_{i = 1}^{N'} \log(\sigma_i^2) - \sum_{i = 1}^{N'} 
	\left(\frac{(Y_i(x_i) - \mu_i(x_i))^2}{\sigma_i^2}\right). \nonumber
\end{align}
In these equations an upper-case $Y(\mathcal{X})$ is used to convey that a
comparison is being made to noisy sample of $Y$-values following the
data-generating mechanism, Observe that score based on full covariance
($\mathrm{score}_{\mathrm{f}}$) is the out-of-sample analog of the log
likelihood (\ref{eq:gpll}), i.e., within an additive and multiplicative
constant of criteria deplored in-sample for GP hyperparameter inference.
Consequently, this criteria is sometimes called the {\em predictive log
likelihood} \citep{gelman2014understanding}.  The ``pointwise'' analog
($\mathrm{score}_{\mathrm{p}}$) offers an approximation for the situation
where full covariances cannot be derived, as sometimes happens with big
testing sets $\mathcal{X}$.  For now, we have full $\Sigma(\mathcal{X})$ for
all three predictors so  Table \ref{tab:toyscores} provides
$\mathrm{score}_{\mathrm{f}}$-values. Notice that by score, the GP-MLE
hyperparameterization is better, confirming our intuition that the PIs for
K-EB are too narrow. Appendix \ref{sec:meuse} shows a similar out-of-sample 
validation exercise comparing kriging to GPs on real data using the Meuse 
river data \citep{Pebesma2009} which is a common 2D kriging data set.

\section{Large-scale modeling via localization}
\label{sec:methods}

Likelihood-based GP inference for hyperparameters
(\ref{eq:gpll}--\ref{eq:other_deriv}) and prediction (\ref{eq:pred}) requires
matrix decomposition for the inverse and determinant of the covariance
structure.  Most kernels, e.g., Gaussian and Mat\'ern, produce dense $K_N$
(and thus $\Sigma_N$), which require $\mathcal{O}(N^3)$ decomposition, say via
Cholesky (which furnishes both inverse and determinant).  This is prohibitive
for $N$ larger than a few thousand.  For example, decomposing a single $40,000
\times 40,000$ matrix on a workstation using 8 cores and specialized linear
algebra libraries (Intel MKL) takes about 10 minutes.  Numerical
optimization of hyperparameters might require hundreds of such decompositions
in search of the MLE via BFGS. With cubic scaling for larger $N$, computation
time quickly explodes to hours or days. $\mathcal{O}(N^2)$ storage of the $N
\times N$ matrix can also become problematic even on the most powerful
workstations.  Kriging-based inference for hyperparameters via variography
bypasses the need to work with an $N \times N$ covariance matrix by binning
the data.  However, Eq.~(\ref{eq:pred}) still requires a dense $N \times N$
inverse to furnish predictions, which is still cubic in computational order.

In the modern age, datasets can easily push into the multimillions and the
methods described in Section \ref{sec:review} quickly become infeasible. For
that reason, there are increasingly many approaches that seek a thrifty
approximation to GP/kriging models. \cite{Heaton2019} give a thorough comparison
of about a dozen recently introduced spatial methods equipped to
handle large data. Here we focus on three representative approaches as a means
of spanning myriad alternatives in a mining context: Ordinary kriging
\citep[OK;][]{Matheron1971} is the standard method in mining/geostats which
makes local (approximate) prediction after full-data variogram-based
hyperparameter estimation; Local Approximate Gaussian Processes
\citep[LAGP;][]{Gramacy2015} can be seen as a likelihood-based contemporary
analog of OK developed in the surrogate modeling community, making it a
natural comparator to OK; finally, the scaled Vecchia approximation (SVecchia)
\cite{Katzfuss2021} uses a global approximation to estimate the full
covariance structure based on similar locality principles as LAGP/OK.  Details
for each of these follow in subsections below.  

Lastly as a baseline, we consider subset GPs trained via a randomly selected, 
computationally feasible, $m \ll N$-sized subset of the data points and use 
them to form an approximation to the full model. Specifically, we build $(X_m, 
Y_m) \subset (X_N, Y_N)$, using the likelihood for hyperparameter inference via 
Eqs.(\ref{eq:gpll}--\ref{eq:other_deriv}) using $(X_m, Y_m)$ and prediction 
similarly following (\ref{eq:pred}). We consider $m$ ranging from 1,000 to 
8,000; we show later in Figure \ref{fig:AGA} that $m > 8{,}000$ is very slow 
and not competitive with the other methods in terms of accuracy out-of-sample.

\subsection{Transductive Modeling}
\label{sec:transductive}

Perhaps the most common solution to big-data matrix issues when predicting via
kriging is to deploy what is known as {\em ordinary kriging}
\citep[OK;][]{Matheron1971}.  OK involves using full-data variography to learn
kernel hyperparameters, and then a {\em local} application of that learned
kernel through predictive equations (\ref{eq:pred}) conditioned only on a
small subset of the data nearby the predictive location(s) of interest.  Let
$x \in \mathcal{X}$ denote the coordinates of one such location, and $X_m(x)
\subseteq X_N$ denote the $m$ ``closest'' (e.g., via Euclidean distance)
members of $X_N$ to $x$, and let $Y_m(x)$ be the $m$-associated output values.
These are sometimes called the $m$-{\em nearest neighbors} (NN) to $x$ in
$X_N$. Then simply apply Eq.~(\ref{eq:pred}) with $(X_m(x), Y_m(x))$ rather
than $(X_N, Y_N)$. When $N$ is so large that the requisite $N \times N$ matrix
decompositions are intractable, choosing $m \ll N$ like $m=50$ can represent a
thrifty-yet-accurate alternative acknowledging that the discarded points $X_N
\setminus X_m(x)$ have vanishingly small impact on the predictive equations
especially when kernels involve exponential decay.

If a multitude of $x \in \mathcal{X}$ are of interest, these may be processed
in serial or, as is increasingly common with modern computing architectures,
in parallel on multiple cores of a workstation and/or nodes of a
supercomputer. Vast predictive grids $\mathcal{X}$ can be processed
efficiently in this manner.  There are several variations on this theme, many
involving how the ``neighborhood'' $X_m(x)$ and its size $m$ are defined. For
example, one may work with a radius $r$ instead, implicitly defining $m$
depending on the local nature of design locations $X_N$ nearby $x$. Suitable
$r$ from a modeling perspective may be selected by the estimated range $R$ of
the semivariogram. However, this does not guarantee a suitably-sized $m$ for
all $x$. One may end up with too small of a neighborhood to make
computationally stable calculations/reliable low-variance predictions, or too
large of one to be carried out efficiently from a computational perspective.
Consequently, there are many hybrids that are often deployed in this space 
\citep[Chapters 11-13]{Wackernagel2003}.

The idea of tailoring a statistical calculation to a predictive task, using
different data and possibly different calculations depending on the predictive
location $x$ of interest, is now known as {\em transductive learning}
\citep{Vapnik2013}.  The transductive moniker is meant to contrast with the
more typical {\em inductive learning} setup where one trains first and
predicts second.  Under transductive learning, the training happens bespoke to
each $x \in \mathcal{X}$, and usually on-demand/in real time.  Examples span
the gamut of statistical modeling enterprises, often offering both speed and
accuracy gains over the inductive analog. Reviewing these would be a
distraction here. Instead, we note that OK is an example of transductive
learning ahead of its time, albeit a somewhat limited one.  Hyperparameter
learning with OK is inductive, whereas posterior predictive conditioning is
transductive.  It is this latter stage where the nonparametric flexibility
really comes from, although one might wonder whether things could improve by
enhancing the degree transductively, as it were.

A prime example of transductive GP modeling from the wider statistical
learning literature is the {\em local
approximate Gaussian process} \citep[LAGP;][]{Gramacy2015}. LAGP is similar to
OK, using $X_m(x)
\subset X_N$ and $Y_m(x)$ analogously for prediction, but it is different in
that it extends the notion of locality to hyperparameter inference via the
(local) likelihood.  That is, the entire process conditions only on $(X_m(x),
Y_m(x))$ for inference (\ref{eq:gpll}--\ref{eq:other_deriv}) and prediction
(\ref{eq:pred}). All inference is off-loaded to numerical optimization. When
the response surface is nonstationary, e.g., benefiting by longer lengthscales
for some $x$ and shorter for others, LAGP offers enhanced reactivity compared
to single, global setting of hyperparameters. Any variation tailored to the
full-GP is easy to port to the local setting because LAGP is just many small
GPs.  Hybrids are possible too.  One such example alluded to earlier involves
pre-scaling or rotating/projecting
\citep[e.g.,][]{wycoff2021sensitivity,sun2019emulating} to handle
non-axis-aligned anisotropy.  However, the biggest difference between LAGP and
OK does not lie in potential for extension; it is how the neighborhood is
defined.

Given fixed $m$, usually chosen via computational considerations (a common
default is $m=50$), it has been known for sometime that the $m$-NNs in $X_N$
to $x$, whether via Euclidean distance or otherwise, do not comprise of an
optimal conditioning set $(X_m(x), Y_m(x))$ under any reasonable criteria
\citep[e.g.,][]{stein2004approximating}.  Example criteria include (Fisher)
information about unknown hyperparameters or, as is usually more relevant when
predictive accuracy is concerned, predictive uncertainty (a.k.a., mean-squared
prediction error).  We note that this, in turn, means that the OK predictor is
also sub-optimal as a transductive learner. However searching for the best
conditioning set $(X_m(x), Y_m(x))$, again under almost any criteria,
represents a computationally daunting task because there are ${N \choose m}$ 
alternatives to explore.

Here, another modern statistical learning idea comes in handy:  {\em active
learning} (AL).  AL is a branch of reinforcement learning/optimal control, or
may be viewed as a modernization of old-fashioned sequential design of
experiments.  In the AL literature, one can often show that a
one-step-at-a-time, {\em greedy} selection of training data is nearly as good
as an exhaustive optimization of some criteria by demonstrating that the
criteria has a {\em submodularity} property \citep{wei2015submodularity}. For
example, it can be shown that repeatedly acquiring training data $(x_{n+1},
y_{n+1})$ such that $x_i$ maximizes the predictive variance $ x_{n+1} =
\mathrm{argmax}_{x} \; \sigma_n^2(x)$ of a GP (\ref{eq:pred}) or neural
network model training only on  $\{(x_i, y_i)\}_{i=1}^n$ obtained previously
(e.g., via similar greedy optimization), well-approximates a so-called maximum
entropy design, i.e., maximizing Shannon information about unknown
hyperparameters (GP lengthscales) for the entire selection $i=1,\dots,N$, say.
This idea is due to \citet{mackay1992information} for neural networks, and
dubbed ALM by \citet{seo2000gaussian} in extension to GPs.

Intuitively, selecting points which have maximum variance will result in a
space-filling design because variance is higher away from the training data.
Also, intuitively spreading points out will increase accuracy and reduce
uncertainty throughout the input space.  But this is coincidental.
Guaranteeing reduced predictive variance everywhere, or at a particular
location (for choosing an LAGP neighborhood), requires a criteria that
squarely targets reduced variance in the region of interest.  A common choice
is integrated mean-squared predictive error \citep[IMSPE;][]{Sacks1989}:
\begin{equation}
	\label{eq:imspe}
	\text{IMSPE}(x_{n + 1}) = \int_{\mathcal{X}} \sigma_{n + 1}^2(x) \; dx
	\approx  \sum_{x \in X_{\mathrm{ref}}} \sigma_{n + 1}^2(x) =  \text{ALC}(x_{n + 1}, X_{\mathrm{ref}}).
\end{equation}
The integral is usually taken over the entire input space, but
$\mathcal{X}$ could be any set. This could be interpreted as a criteria for
the entire design  $X_{n + 1} = [ X_n ; x_{n + 1}]^\top$, or simply to select
the next input $x_{n+1}$ in an active learning context: $x_{n + 1} =
\mathrm{argmin}_{x \in \mathcal{X}} \ \text{IMSPE}(x)$. Due to submodulatirty,
both (approximately) optimize the IMSPE criteria over all $X_{n+1}$.  
\citet{Cohn1996} developed this for neural networks, and \citet{seo2000gaussian} extended it to GPs. Notice
that ALC approximates the integral as a quadrature over a discrete
reference set $X_{\mathrm{ref}}$.  This is not necessary for GPs, because the
integral is analytic when following Eq.~(\ref{eq:pred}), but it is for neural
networks.

For LAGP, the goal is to get as accurate of a prediction at $x$ as 
possible, which can be interpreted as a singleton $\{x\} = \mathcal{X} = 
X_{\mathrm{ref}}$, effectively discarding the sum or integral.  We can select a 
new $x_{n + 1} = \mathrm{argmin}_{x_{n + 1}} \; \text{ALC}(x_{n + 1}, x)$, and 
repeated applications will approximate a ``local'' optimal design for 
predicting at $x$. This would usually be applied for selecting new training 
data in an AL context, but for LAGP we already have a fixed training data set 
$(X_N, Y_N)$ and so we desire a subsample instead:  $x_{n + 1} = 
\mathrm{argmin}_{x_{n + 1} \in X_N \setminus X_n} \; \text{ALC}(x_{n + 1}, x)$, 
which is even easier than a continuous search potentially everywhere in the 
input space.

\begin{figure}[ht!]
	\centering
	\includegraphics[height=7cm,trim=20 35 20 50]{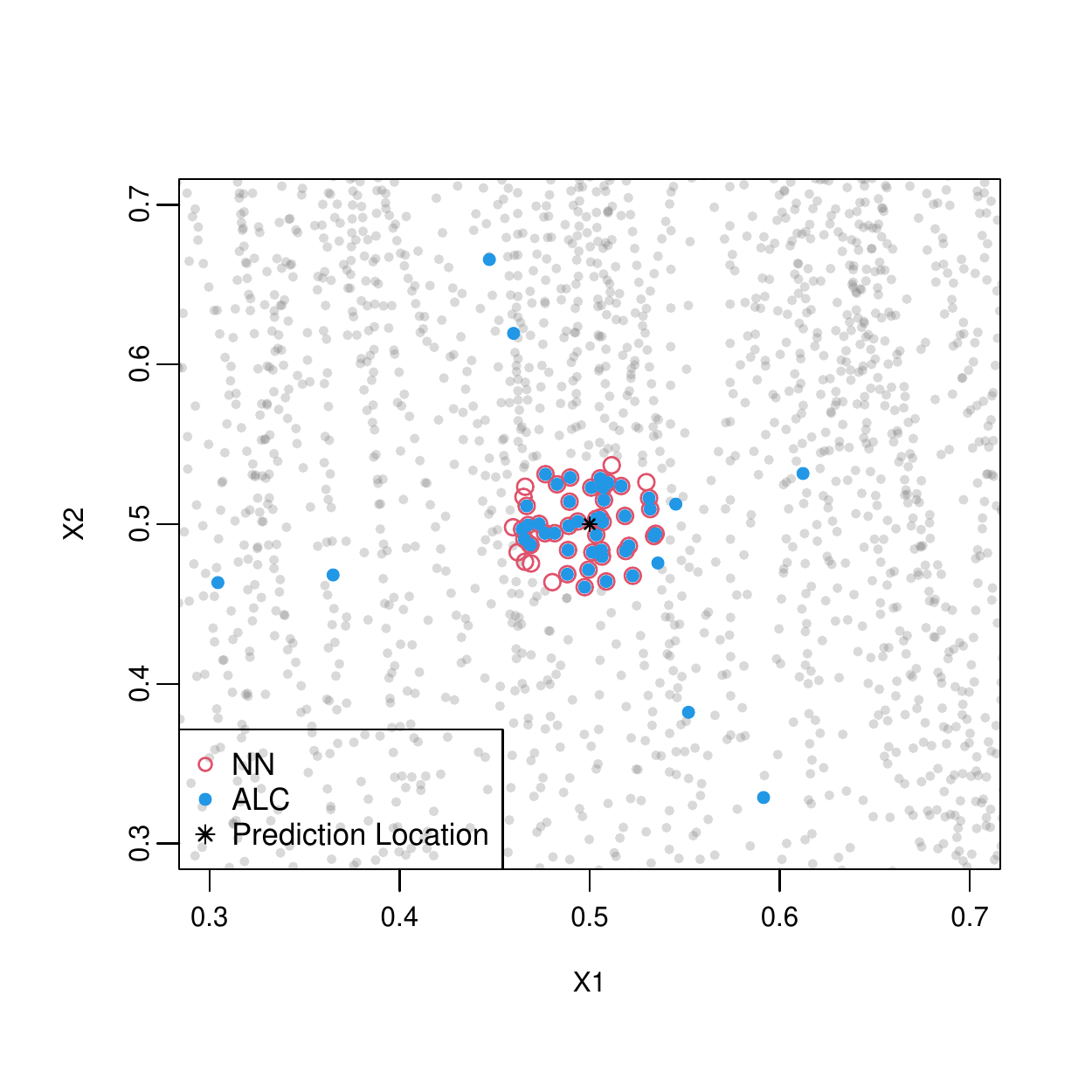}
	\caption{ALC neighborhood for $n=50$ on simulated borehole data.}
	\label{fig:laGP}
\end{figure}

In practice, early ALC acquisitions (small $n$) result in neighborhoods that
are indistinguishable from NN. However, later acquisitions (larger $n$), after
many NNs have been conditioned upon near $x$, the local ALC/IMSPE criteria
prefers acquiring ``satellite'' points farther away.  Intuitively, information
farther afield becomes more valuable as NNs accumulate near $x$: you want
$x_{n+1}$ to be both close to $x$ but far from $X_n$.  At the start the former
dominates, but eventually the latter has higher weight in the criteria. The
end result for $n=50$, seen in Figure \ref{fig:laGP}, involves ten or so
satellite points, departing from an OK or NN subdesign set of the same size.

To a certain extent, LAGP has a ``chicken or the egg'' problem when dealing with
anisotropy. Neighborhoods selected for $x$ are based on Euclidean distance to
determine hyperparamters, like $\hat{\theta}_k(x)$, but those values control
notions of distance differently in each input coordinate through the
kernel.  OK experiences this problem too, albeit to a lesser extent when
anisotropy is handled by the practitioner as a pre-processing step.  Several
remedies have been proposed. The original LAGP paper \citep{Gramacy2015}
suggested initializing with a default, isotropic $\theta_0$ for all testing
locations $x$, upon which neighborhoods are built (e.g., via ALC) and local,
anisotropic local $\hat{\theta}(x)$ are learned through local MLEs separately
for each $x$.  This can then be repeated, with neighborhoods based on those
$\hat{\theta}(x)$, until things stabilize. Subsequently, a simpler/better
approach was promoted by \cite{sun2019emulating} that is more akin to OK
pre-processing, but still completely automated. \cite{Liu2015} showed that
unbiased (global) lengthscales $\hat{\theta} = (\hat{\theta}_1,\dots,
\hat{\theta}_d)$ can be estimated via MLEs from carefully constructed data
subsets of large $X_N$ without expense cubic in $N$.  Once these have been
learned, they can be used to pre-scale inputs $X$ as $X_k / \sqrt{\theta_k}$
so that the implied MLE global lengthscale is $\hat{\theta}_0 = 1$ under
squared-distance kernels like in Eq.~(\ref{eq:kernels}).  In this transformed
space,  Euclidean distance can be used to determine neighborhoods.  This
pre-scaled LAGP has become the default setup, 
and has since been
extended to other input ``warpings'' by \citet{wycoff2021sensitivity}.

Taken as a predictive field over a densely gridded testing set of $x$-values,
both OK and LAGP (NN or ALC) are discontinuous. Prediction at each point $x
\in \mathcal{X}$ is processed independently, both in a statistical and
computation sense.  So two testing locations right next to each other
may have substantively different predictions (in mean and/or variance) because
different neighborhoods are used for conditioning. This can
be an asset when the response surface exhibits regime/abrupt changes.
In tamer, more stationary settings, a non-smooth prediction could be
detrimental to statistically efficient and aesthetically pleasing analysis.

\subsection{The Scaled Vecchia Approximation}
\label{sec:svecchia}

The Vecchia GP approximation \citep{Vecchia1988} borrows the neighborhood idea
while providing a global model for smooth predictions. It relies on
a familiar identity for joint distributions:
\begin{align}
	\label{eq:vecchia}
	p(y) = p(y_1) p(y_2 | y_1) \dots p(y_n | y_1, y_2, y_3, \dots y_{n - 1})
	&= \prod_{i = 1}^N p(y_i | y_{k(i)}) &&\text{where} \quad k(i) = 
	\{j : j < i\} \\
	&\approx \prod_{i = 1}^N p(y_i | y_{c(i)}) &&\text{where} \quad 	c(i) \subset k(i)	\nonumber
\end{align}
The first line above (equality) is true for any re-indexing of the variables
$y = y_1,\dots,y_n$, and for any $y$ -- not specifically for GPs. The
approximation (second line) arises from dropping some of those conditioning
variables. Let $m$ denote the maximum size of those sets, i.e., so that $\vert
c(i) \vert = \min(i - 1, m)$, controlling the fidelity of the approximation --
more severely for $m \ll i$.  The quality of this approximation is determined
by the indexing (i.e., the ordering of the conditionals), size $m$, and which
of the conditioning variables $k(i)$ are dropped in $c(i)$ when $m < i$.

Specifically for GPs, one may view the likelihood (\ref{eq:gpll}),
in this context:
\begin{align}
	\label{eq:vecl}
	L(\phi; Y_N) &\approx \prod_{i = 1}^N L(\phi; y_i | 
	y_{c(i)})\\ 
	&= (2\pi)^{-\frac{N}{2}} \left(\prod_{i = 1}^N 
	\sigma_i^2\right)^{-\frac{1}{2}} 
	\exp\left\{-\sum_{i = 1}^N\frac{1}{2\sigma_i^2} (y_i - \Sigma(x_i, 
	X_{c(i)})\Sigma(X_{c(i)}, X_{c(i)})^{-1} y_{c(i)})^2\right\} \nonumber
\end{align}
where $\Sigma(\cdot, \cdot)$ is defined as in Section \ref{sec:GPs} and
$\sigma_i^2 = \Sigma(x_i, x_i) - \Sigma(x_i, X_{c(i)}) \Sigma(X_{c(i)},
X_{c(i)})^{-1}\Sigma(X_{c(i)}, x_i)$ is the predictive variance at location
$i$ given the conditioning set, $c(i)$. Since distance in the input space, via 
$\Sigma(\cdot, \cdot)$ is fundamental to GP inference and prediction, one can 
think of $c(i)$ as defining a ``neighborhood.'' In that context it makes sense 
(as it did for OK and LAGP) to include in the neighborhood those indices whose 
input values $X_{c(i)}$ are closer to $x_i$. 

Eq.~(\ref{eq:vecl}) is similar to (\ref{eq:gpll}) except instead of performing
one $N \times N$ matrix decomposition (for inverse and determinant) in
$\mathcal{O}(N^3)$ time, the Vecchia approximation involves $N$ smaller inversions of $m
\times m$ matrices, requiring $\mathcal{O}(Nm^3)$ flops. If $m$ is small, typically 
between 10 and 25 \citep{Datta2016, Katzfuss2021}, $\mathcal{O}(Nm^3)$ is quasilinear in
$N$ \citep{Katzfuss2020}. Further computational gains can be realized through
sparse-matrix libraries and parallelization by re-writing the likelihood
through a Cholesky decomposition of the precision matrix of $Y_N$, denoted as
$U$:
\begin{align}
	L(\phi; Y_N) &\approx (2\pi)^{-\frac{N}{2}} \left(\prod_{i = 1}^N 
	\sigma_i^2\right)^{-\frac{1}{2}} 
	\exp\left\{-\frac{1}{2}\sum_{i = 1}^N (Y_N^\top U_i U_i^\top Y_N)\right\}
	\label{eq:veclchol}\\
	&= (2\pi)^{-\frac{N}{2}} |U U^\top|^{\frac{1}{2}} 
	\exp\left\{-\frac{1}{2}(Y_N^\top U U^\top Y_N)\right\},
	\nonumber
\end{align}
where  $U_i$ is a $1 \times N$ vector whose $j^\mathrm{th}$ entry is:
\begin{equation}
	U_i^{(j)} =
	\begin{cases}
		\frac{1}{\sigma_i} & i = j\\
		-\frac{1}{\sigma_i} \left(\Sigma(x_i, x_j)
		\left(\Sigma(X_{c(i)}, X_{c(i)})^{-1}\right)^{(j, j)}\right) & j \in 
		c(i)\\
		0 & \ \text{otherwise}.
	\end{cases}
	\label{eq:udef}
\end{equation}
Observe that there are no $N \times N$ matrix decompositions involved, and
that any inverses are implicit in the Cholesky factor $U U^\top$, which is
sparse.

One may maximize the likelihood (\ref{eq:veclchol}) to estimate
hyperparameters \cite{guinness2018permutation}.  Prediction follows the 
classical setup
(\ref{eq:pred}), forming $Y(\mathcal{X}) \mid Y_N$ by stacking training and
testing responses:
\[
\begin{bmatrix} Y_N \\ Y(\mathcal{X}) \end{bmatrix} \sim
\mathcal{N}_{N + N'} \left(
\begin{bmatrix}
	0 \\ 0 
\end{bmatrix}, 
\begin{bmatrix}
	UU^\top = \sum_{i = 1}^N U_iU_i^\top & \sum_{i = 1}^N U_i \sum_{i = 
		1}^{N'} U_i^{\prime\top} \\
	\sum_{i = 1}^{N'} U'_i \sum_{i = 1}^N U_i^\top & U'U^{\prime\top} = 
	\sum_{i 
		= 1}^{N'} U'_iU_i^{\prime\top}
\end{bmatrix}^{-1}
\right).
\]
Here we have introduced a new $N' \times N'$ matrix, $U'$ following
Eq.~(\ref{eq:udef}), via $\mathcal{X}$ rather than $X_N$. Then, the analog of
Eq.~(\ref{eq:pred}) yields
\begin{equation}
	\label{eq:svecchiapred}
	\mu_N(\mathcal{X}) = -(U' U^{\prime\top})^{-1 }\sum_{i = 1}^{N'} U'_i 
	\sum_{i = 1}^N U_i^\top Y_N \quad \mbox{ and } \quad
	\Sigma_N(\mathcal{X}) = U'U^{\prime\top}.
\end{equation}
So the Vecchia GP approximation provides a full joint distribution. Moreover,
a single prediction ($\mathcal{X} \equiv \{x\}$), after training, requires just
$\mathcal{O}(m^3)$ additional time \citep{Katzfuss2021}, assuming cached
values of $U$. A total of $\mathcal{O}((N' + N)m^3)$ flops are required for
inference and prediction, as opposed to $\mathcal{O}(N'^3 + N^3)$ for an
ordinary GP.

All that remains is to determine the ordering of indices $i$ in $y_i$ and the
composition of the neighborhood sets $c(i)$, since not all choices (when $m
\ll n$) lead to equally good approximations (\ref{eq:vecchia}--\ref{eq:vecl}).
One option is to follow the LAGP playbook and attempt to optimize over these
variables.  However this has proved elusive in the literature because an
exhaustive search over alternatives would be combinatorially cumbersome, and
there is no obvious greedy approach that enjoys submodularity for active
learning.  Nevertheless there are rules of thumb that make sense intuitively.
Many orderings work well
\citep{stein2004approximating,guinness2018permutation,katzfuss2021general},
but there is a consensus in the literature
\citep{stroud2017bayesian,Datta2016,wu2022variational} for random indexing.
Likewise, those authors prefer  NN conditioning sets $c(i)$ comprised of
indices $j < i$ whose $x_j$-values are closest to $x_i$.  This choice has been
dubbed NNGP by \citet{Datta2016}, although it is important to note that NN are
not being used in the same way as LAGP or OK.

Since distances are involved in NN calculations, the Vecchia approximation
faces the same ``chicken or the egg'' problem as LAGP in the face of
anisotropy. To help, \cite{Katzfuss2021} describe a scheme similar to
pre-scaling for LAGP which updates lengthscales $\hat{\theta}_k$ via Fisher
scoring \citep{Osborne1992}, then re-scales inputs so that NNs can be
recalculated, and repeats.  \cite{Katzfuss2021} call this ``scaled Vecchia''
(SVecchia), and argue that it works best with a maximin \citep{Johnson1990}
indexing.  We adopt SVecchia as our preferred variation on this theme, in part
because it is neatly packaged in software [Section \ref{sec:results}].

\begin{figure}[ht!]
	\centering
	\includegraphics[height=5cm,trim=10 35 10 60]{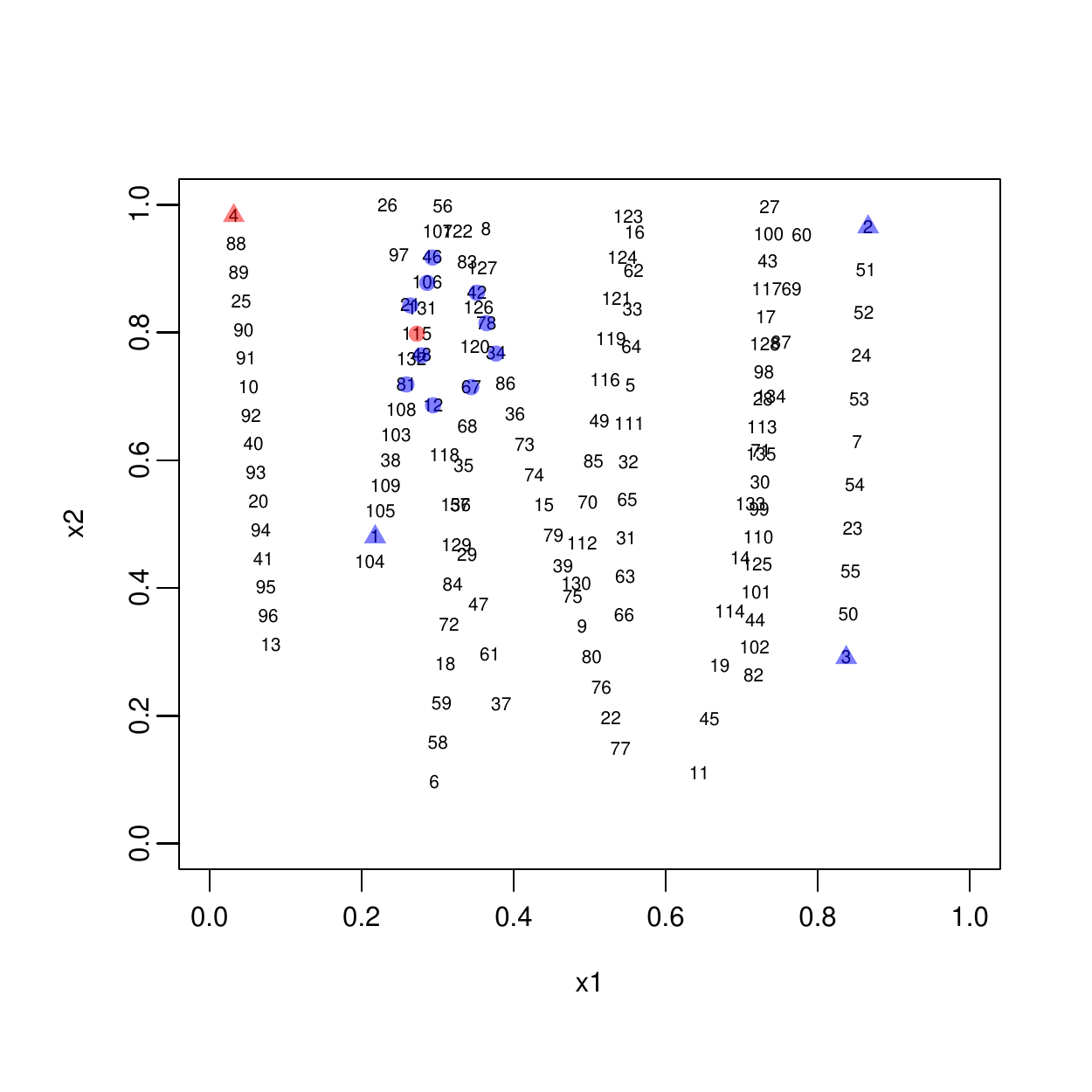}
	\includegraphics[height=5cm,trim=10 35 10 60]{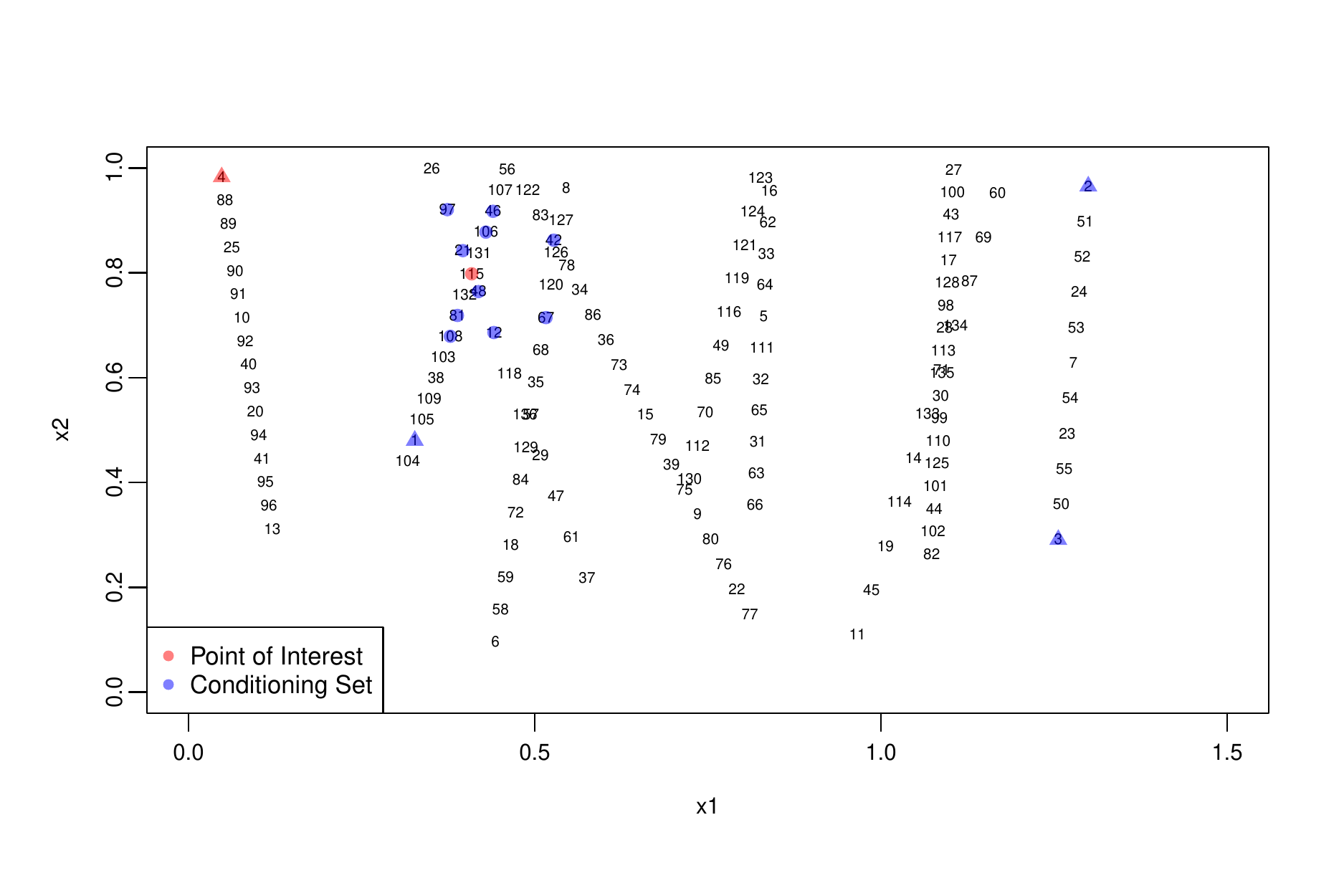}
	\caption{Conditioning sets for two inputs using $m = 10$: point 4 
	(triangles) and 115 (circles). Left shows the original space in coded
	inputs, and right shows $x_1$ pre-scaled by $1/\sqrt{\theta_1} = 2/3$.}
	\label{fig:SVecchia}
\end{figure}

Figure \ref{fig:SVecchia} shows an illustration using simulated borehole
data, providing conditioning sets of size $m = 10$ for two points, labeled
with indices $i=4$ (triangles) and $i=115$ (circles) respectively. The left
plot shows what the conditioning sets look like in the raw, unscaled space
while the right is after scaling $x_1$ by $\frac{2}{3}$; both using the same
maximin ordering for easy comparison. The scaling has no effect on the small
indices, since point 4, for example, can only condition on $k(4) = c(4) =
\{1,2,3\}$. See that the lower indices in $c(4)$ (purple triangles) are spread
out through space due to maximin ordering. However, point 115 has $\vert
k(115) \vert = 114$ points to choose from for it's neighborhood $c(115)$.
Consequently, the conditioning sets are different between the unscaled and
scaled versions. Observe that point 97 is closer than point 34 in the scaled
plot, but farther in the unscaled plot. 

\section{Empirical evaluation on ore data} 
\label{sec:results}

Here we shall expand on our out-of-sample analysis from Section \ref{sec:oosv}
in order to illustrate how modern approximate GP and kriging alternatives
[Section \ref{sec:methods}] compare on two, real and large-scale ore data
sets. The layout is as follows. We first introduce the data, evaluation
metrics, and detail software in Section \ref{sec:implement}.  Section
\ref{sec:cv} presents our validation apparatus, which offers a subtle twist on
conventional methods to respect the borehole nature of data
acquisition/measurement.  Here we also provide results from  our first sets of
comparisons.  Finally, Section \ref{sec:imp} presents one of several
potentially more nuanced analyses that, we believe, is only possible (with
ease) in the fully probabilistic GP setting: coping with left-censoring
prevalent in ore measurements.

\subsection{Implementation details}
\label{sec:implement}

Our ore data involve three-dimensional inputs, indicated as longitude,
latitude and depth in standard units.  Although these data record measurements
(potential ore outputs) for multiple elements, our analysis here focuses on
(log) gold concentration in parts per million. The two data sets are for geographically disparate mining sites that are characterized
by different ore forming processes. The first one records more than
$150{,}000$ measurements from approximately four-thousand boreholes; the
second has $N \approx 500{,}000$ from $8{,}000$ holes. The second data set
also has a substantial number of left-censored values (i.e., thresholded
measurements below the detection limits of the apparatus used to sample the
core).  For example, about 40\% of the gold measurements in these data are
recorded as 0.05. There are a smaller number of higher limiting
values as well. We shall detail how we handle this with two different
treatments in Sections \ref{sec:cv} and \ref{sec:imp}. We are
deliberately being vague about many aspects of our data in order to honor
confidentiality agreements with mine operators.

We wish to draw an out-of-sample comparison between the methods in Section
\ref{sec:methods} on these data.  In addition to RMSE and proper score
(\ref{eq:rmse}--\ref{eq:score}), we also report time, considering both compute
(machine) time and practitioner (human) time.  Machine time is measured
precisely, in seconds, for execution on an eight-core hyperthreaded Intel Core
i9-9900K CPU at 3.6GHz with 128GB RAM and Intel MKL linear algebra
subroutines.  Human time is more subjective/imprecise, and we shall have more
to say about that in due course. It is worth remarking that none of the
small-data/exact methods from Section \ref{sec:review} are applicable when $N
\gg 10{,}000$, as we have here.  Approximation is essential. One simple option
is to (randomly) subset the data to a manageable size and apply exact
inference on that subset.   We consider variously sized ``subset GPs'' as a benchmark.


For the approximate methods of Section \ref{sec:methods}, we leverage the
following software libraries.  The {\tt laGP} package \citep{Gramacy2016} for
{\sf R} \citep{R2021} on CRAN accommodates subset GP, LAGP, and SLAGP. Its
implementation is primarily in {\sf C}, with {\tt OpenMP} for symmetric
multi-processing (SMP) parallelization. Only Gaussian kernels are supported by
this software.  We use defaults throughout, including ALC neighborhoods of
size $m=50$. An SVecchia implementation is provided by
\cite{Katzfuss2021},\footnote{https://github.com/katzfuss-group/scaledVecchia}
which piggy-backs off of two {\sf R} packages: {\tt GpGp}
\citep{guinness2018permutation} and {\tt GPVecchia} \citep{GPvecchia}. Although
primarily in {\sf R} under-the-hood, {\tt Rccp} \citep{Rcpp} is used for key
subroutines. These, and other calculations based on sparse matrix libraries
\citep{Matrix} for efficient linear algebra, are also SMP parallelized,
although to a lesser degree compared with {\tt laGP}.  Here we use the
Mat\`ern kernel and other defaults throughout, including a conditioning set
size of $m=25$. Conventional pre-processing is used to code spatial inputs to
the unit 3-cube as recommended by both {\tt laGP} or {\tt SVecchia}
documentation.

OK is provided by {\tt GSLIB} \citep{Deutsch1997}, which is in {\sf Fortran}
assisted by a shell scripting interface. Whereas the other methods are
plug-and-play modulo with few choices that are largely relegated to defaults,
interacting with GSLIB is a human-intensive endeavor, involving extensive
pre-processing. For example, the analyst must first generate an experimental
semivariogram (\ref{eq:semivar}) before fitting a positive-definite kernel
function. When anisotropic spatial correlation is suspected, the analyst may
also have to prepare a variogram map to identify the principal directions of
spatial continuity, and then generate experimental semivariograms for two or
three directions. This task is also complicated by numerous choices that are
part of the analysis, e.g., angular tolerances for directional search windows
and different models for spatial autocorrelation (exponential, spherical,
Gaussian, etc.) When the variogram analysis is complete, the analyst may also
have to complete a coordinate axis rotation to align the dataset with the
directions of maximum and minimum spatial correlation, and also experiment
with numerous choices for the OK algorithm itself, e.g., search window size
and shape, minimum/maximum observations to be use for each estimate. When
working with spatially isotropic data, pre-processing may require only an hour
or so of additional work; this time expense increases dramatically
when faced with anisotropic data.


\subsection{Validation exercise}
\label{sec:cv}

As mentioned in Section \ref{sec:oosv}, it 
is important to compare metrics (\ref{eq:rmse} \& \ref{eq:score}) on 
out-of-sample data. Unlike Section \ref{sec:oosv}, when dealing with real data, 
we cannot calculate these metrics on a dense grid spanning the entire input 
space. In practice, out-of-sample validation occurs by training the model on  
90\% of the data and testing on the other 10\%. Appendix \ref{sec:meuse} shows 
this more in depth.
Repeating that randomization multiple times mitigates the so-called Monte Carlo 
(MC) error metrics like RMSE and score. Here we use $K=10$-fold {\em cross 
validation} \citep[CV;][Chapter 7]{Hastie2001} to average over train--test 
partitions while controlling MC error further by ensuring that each data 
element is used exactly once for testing, and complementarily exactly nine 
times for training.  CV commences by first shuffling the data, and then evenly 
dividing it into a partition of $K$ mutually-exclusive {\em folds}, then 
iterating over those folds $k = 1,\dots,10$, forming a testing set of the data 
in the $k^{\mathrm{th}}$ fold while taking the complement as the training set.  
In this way, $K$ metric evaluations (like RMSE, score or time) can be 
calculated and summarized for comparison.

Early attempts at a CV evaluation of the methods in Section \ref{sec:methods},
after this fashion, revealed a shortcoming in the context of our
borehole-driven ore data sets. Namely, the best predictors of a particular
held-out testing element were almost always comprised entirely of members of 
the training data coming from the same borehole. With boreholes ``holding'' 
approximately thirty to sixty data elements each, depending on the hole and the 
data set, this meant that it was highly probable that an accurate prediction 
could be made trivially just by those nearby evaluations.  This conveyed a 
substantial advantage to OK and LAGP.  We determined that it would be more
realistic, and more fair, to hold out entire boreholes for testing, rather
than partitioning the data on individual data elements regardless of which
borehole they were in.  The idea is to simulate what might happen if we were
to predict measurements for a new borehole that has not been drilled yet.

Toward that end, we built a custom CV which randomly partitioned our data into 
$K$ folds of roughly equally-sized boreholes instead. In this way, boreholes 
are
``tested'' all at once, without being able to lean on other data within the
same borehole for training. 
\begin{figure}[ht!]
\centering
\includegraphics[width=8.5cm, trim=20 30 0 40]{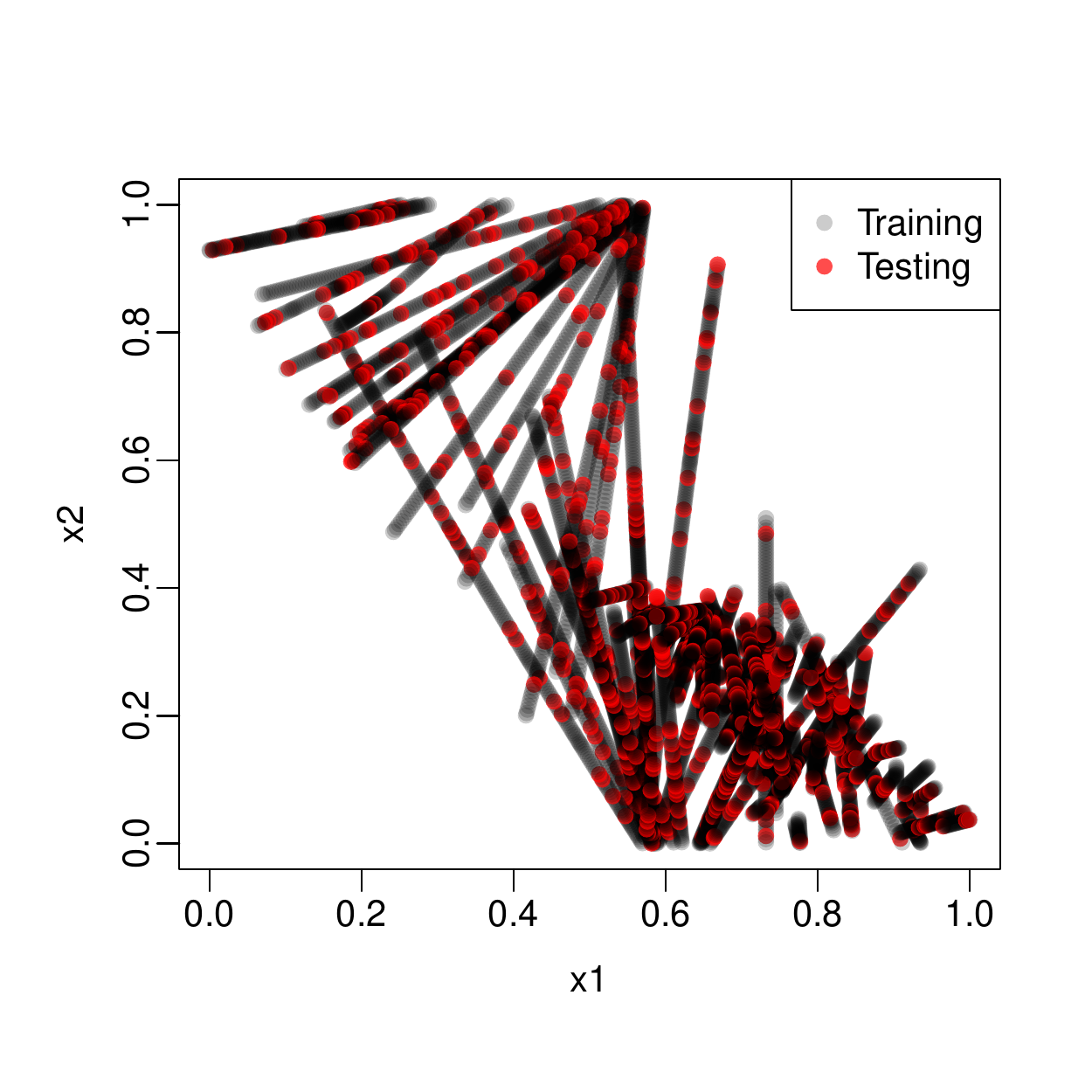}
\includegraphics[width=8.5cm, trim=10 30 0 40]{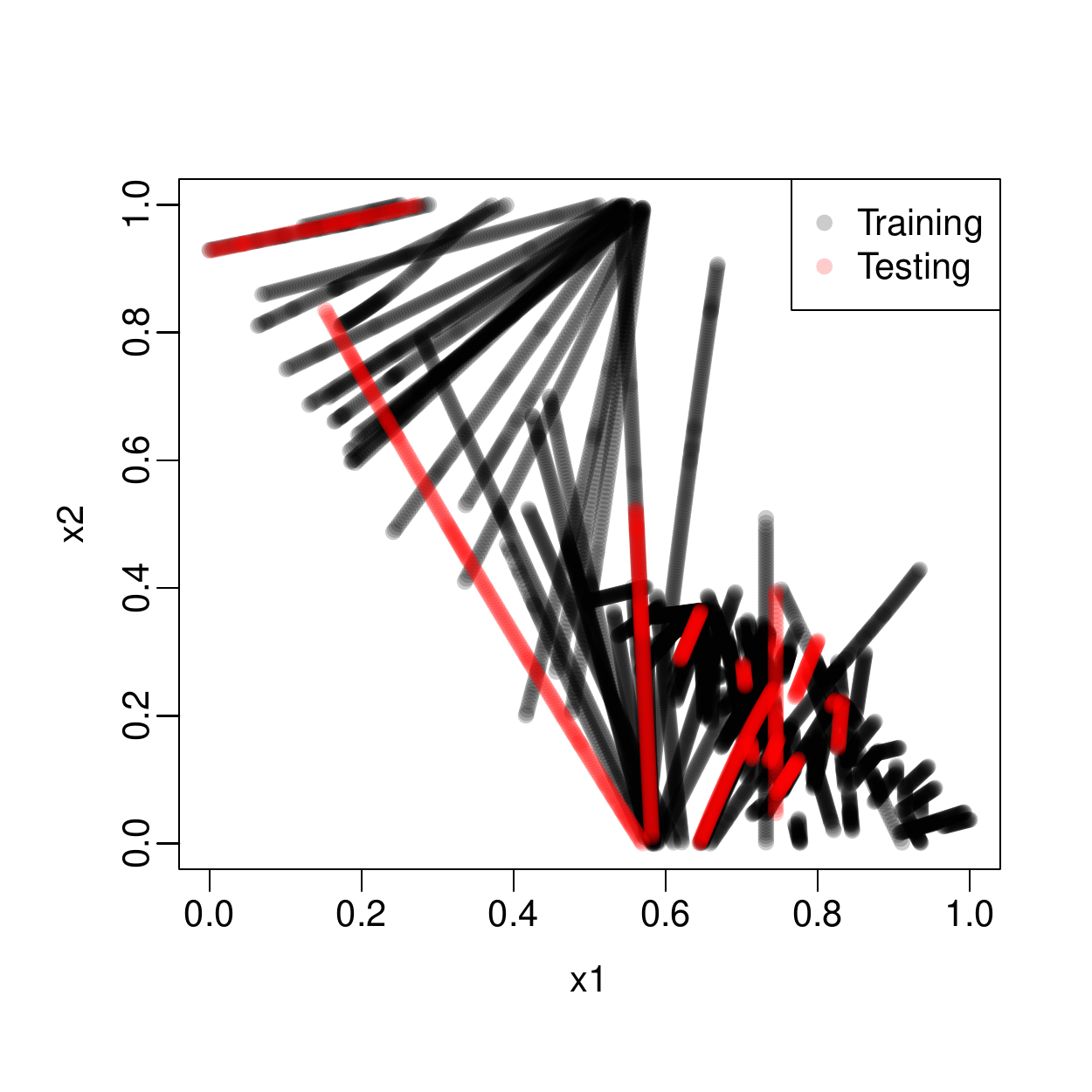}
\caption{Testing and training sets for the 2d project from Figure 
\ref{fig:boreholes} with ordinary CV on the left and borehole-preserving CV 
on 
the right.}
\label{fig:bcv}
\end{figure}
Figure \ref{fig:bcv} shows one such training and testing partition via a
single fold of this ``borehole-preserving CV.'' Finally, it is worth remarking
that all of our comparator methods use exactly the same CV folds.

\subsubsection*{Ore data set one}

\begin{figure}
	\centering
	\includegraphics[width=5.68cm, trim=20 20 0 50]{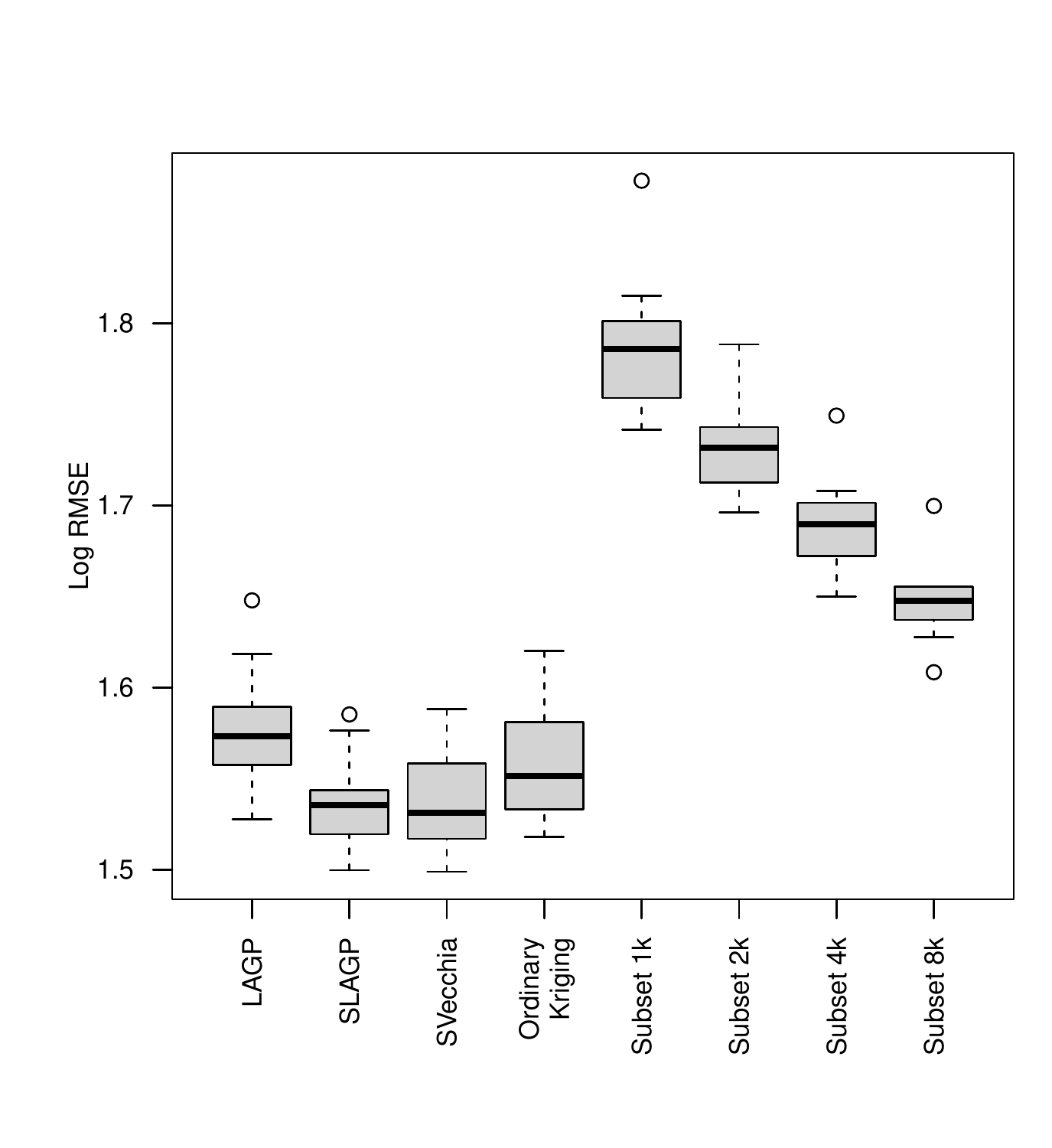}
	\includegraphics[width=5.8cm, trim=10 20 0 50]{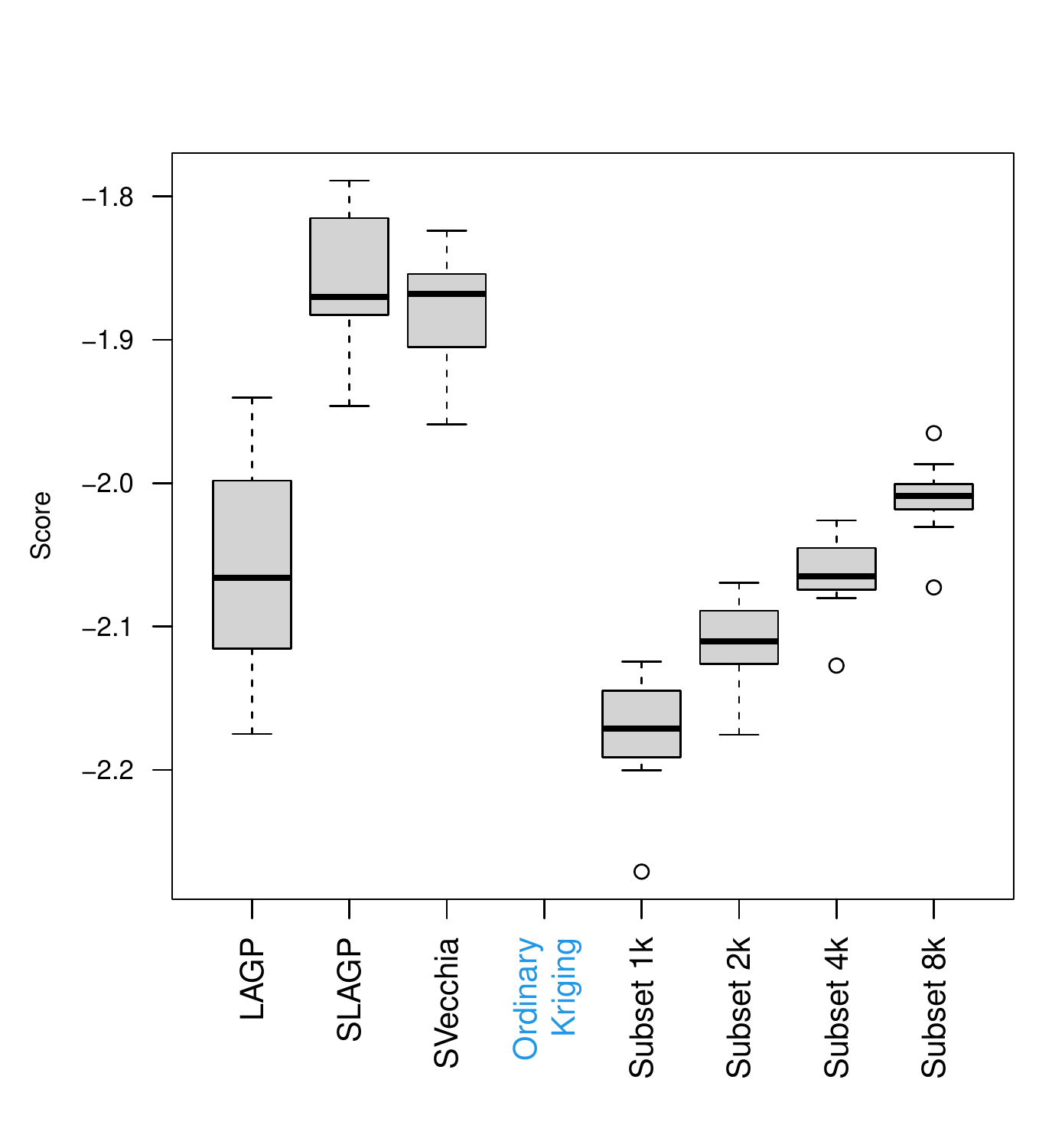}
	\includegraphics[width=5.65cm, trim=10 20 10 50]{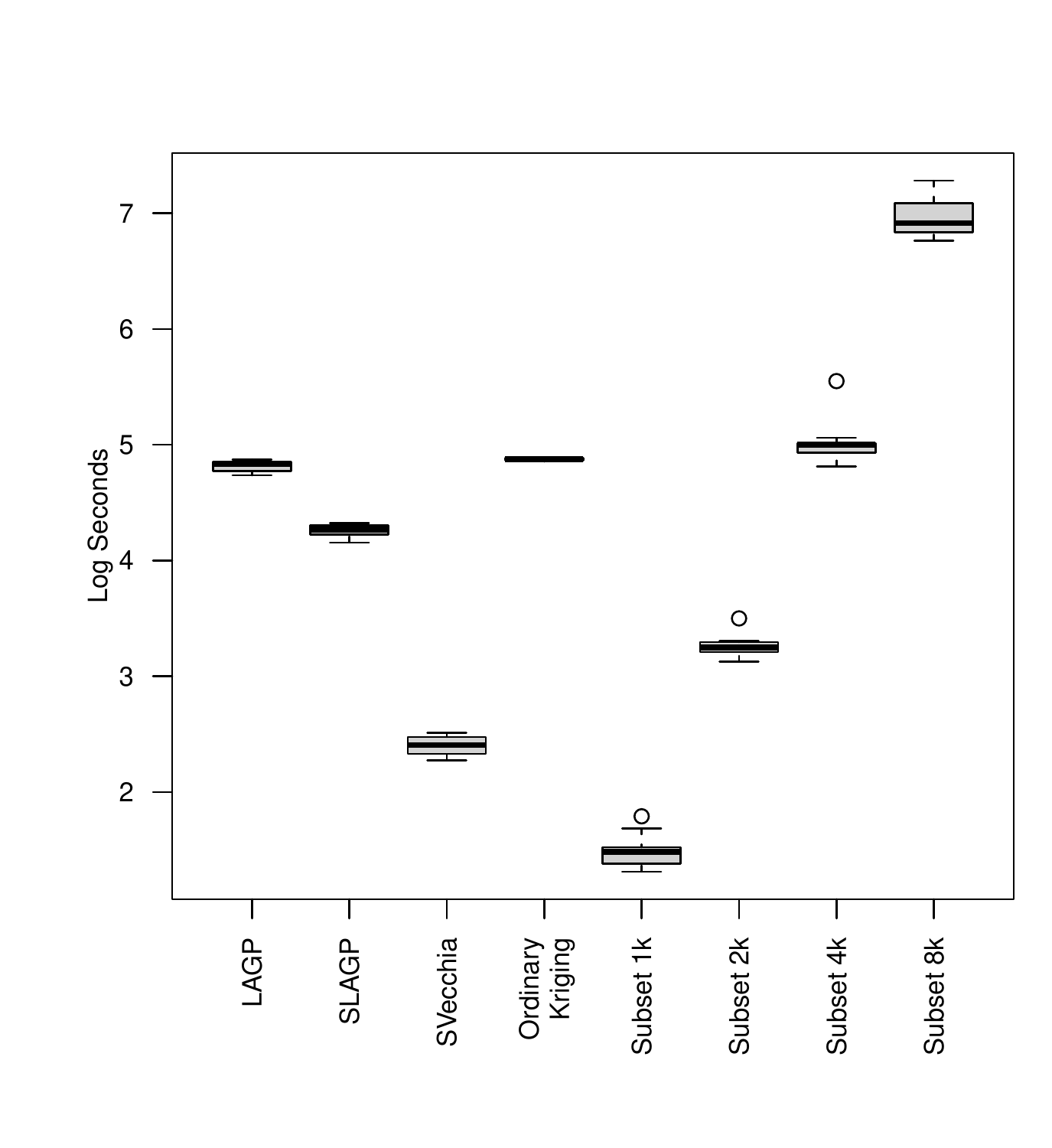}
	\caption{Drillhole-presererving 10-fold CV summary for the 
	first data set. Left: RMSE (smaller is better); Middle: $\mathrm{score}_p$
	(higher is better). Right: compute time (smaller is better).}
	\label{fig:AGA}
\end{figure}

With this setup, Figure \ref{fig:AGA} shows log RMSE (\ref{eq:rmse}), score
($\mathrm{score}_p$ from Eq.~(\ref{eq:score})) and compute time for each of
our methods for the first, smaller data set. The big takeaways are that
SVecchia, OK and SLAGP are all competitive with each other in terms of RMSE;
SLAGP and SVecchia are competitive in score with similar medians. Observe that
SLAGP improves upon ordinary LAGP for both RMSE and score. Score for OK could
not be calculated because GSLIB does not furnish predictive variances. GSLIB
provides standard errors on the mean of the prediction, but those can
substantially under-estimate out-of-sample variance.  
In terms of time, SVecchia takes
seconds and LAGP takes a couple minutes to run, both with essentially zero
``human time.'' We report that OK takes 
several hours
of human time to
perform a variography analysis, choosing between competing kernel formulations
and parameterization and to determine an appropriate rotation and pre-scaling
of the data in order to cope with an otherwise isotropic formulation.  After
that has been done, training and prediction takes about the same amount of
time as (S)LAGP. It is interesting that SLAGP is faster than LAGP despite
involving more algorithmic steps: first fit a global subset model, then local
models on transformed inputs.  The explanation is that, after pre-scaling,
local MLE calculations are easier: they require many fewer iterations to
converge. Computation time for the subset-GP methods explodes with
$\mathcal{O}(m^3)$ flops as $m$ grows.

Our conclusion from this experiment is that, although SLAGP edges out SVecchia
on accuracy and UQ for this problem, SVecchia is slightly better overall
because of its substantially lighter time commitment.  However, for an
individual (or entire borehole) prediction, SLAGP times are orders of
magnitude faster -- the times in the right panel of Figure \ref{fig:AGA} are
for all boreholes in the fold -- because each calculation is independent of
others.  For a one-off prediction it is the clear winner.  
Although raw accuracy is similar compared to OK, the modern GP
methods are hands-off, provide full UQ, and are faster to train/predict.

\subsubsection*{Ore data set two}

A similar analysis for the second, larger data set, is nuanced because of the
substantial left-censoring.  One option is to ignore the censoring and treat
the recorded values as the actual values.  If there were a small number of
such values, sporadically located in the input space, this might work well.
However, there are a sizable number (more than 40\%), and they cluster nearby
one another.  Having some responses smoothly vary in the input space, with
others ``flatlining'' at 0.05, say, for most or all of a borehole represents an
almost pathological contrast to typical smoothness assumptions underlying GP
(and kriging) methods.  So to start with, we removed these values, and dealt
with borehole-preserving CV only on the remaining $259{,}555$ data records. In
Section \ref{sec:imp} we shall discuss an imputation scheme for bringing
these observations back into the folds. The remaining data still contain a
moderate degree of left-censoring which we largely ignore except when an
entire borehole contains the same (thresholded) gold response.  In that
case, we collapse those records into three data points -- two ends and midway
point -- all with the same gold value.  This collapsing is especially
important for LAGP and OK because, due to their local nature, those methods
occasionally have neighborhoods consisting of data from one or two boreholes
only.  If those measurements lack diversity due to thresholding,
training can result in numerical singularities.

\begin{figure}[ht!]
	\centering
	\includegraphics[width=7cm, trim=20 40 0	50]{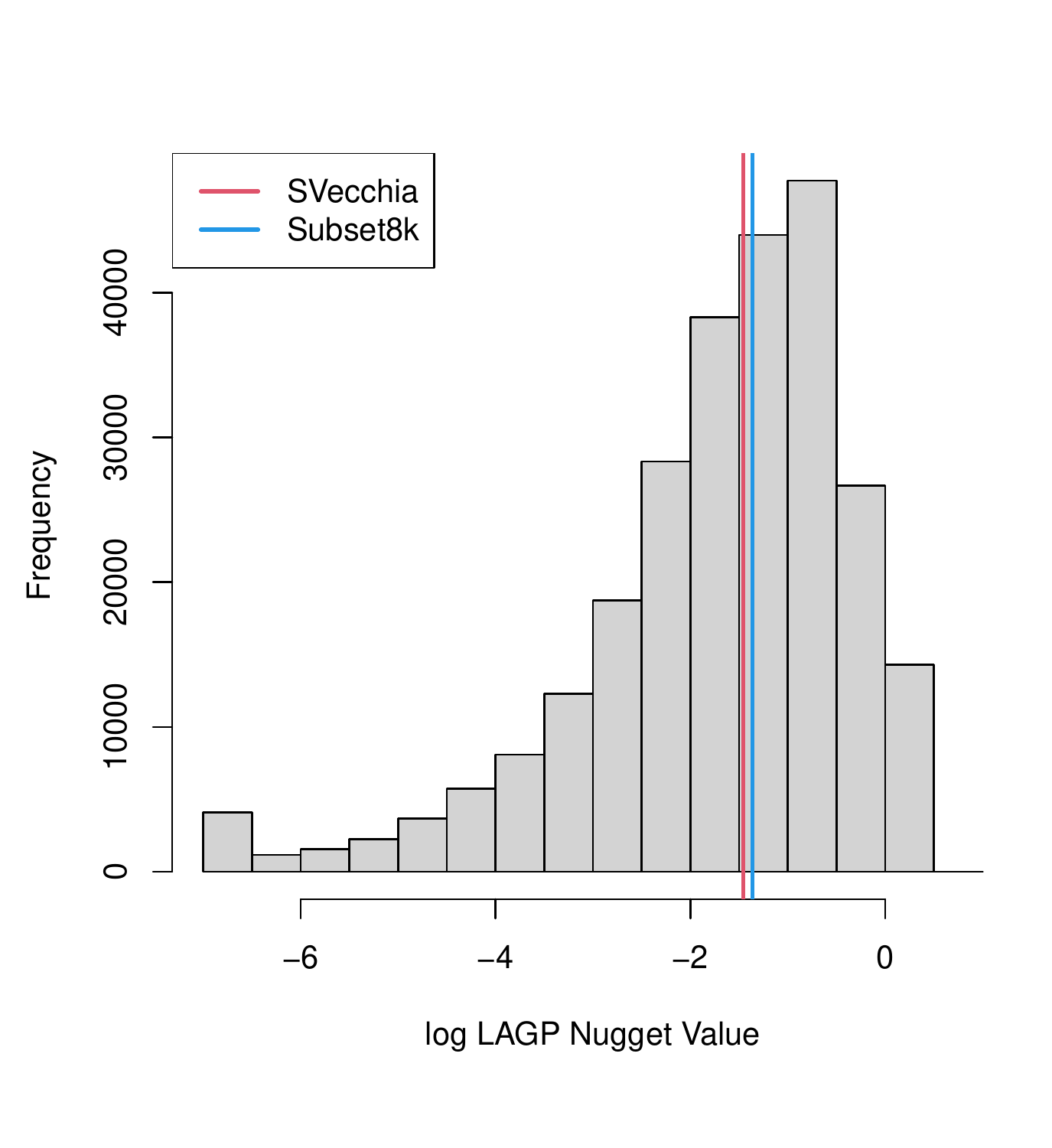}
	\includegraphics[width=9cm, trim=0 20 0 10]{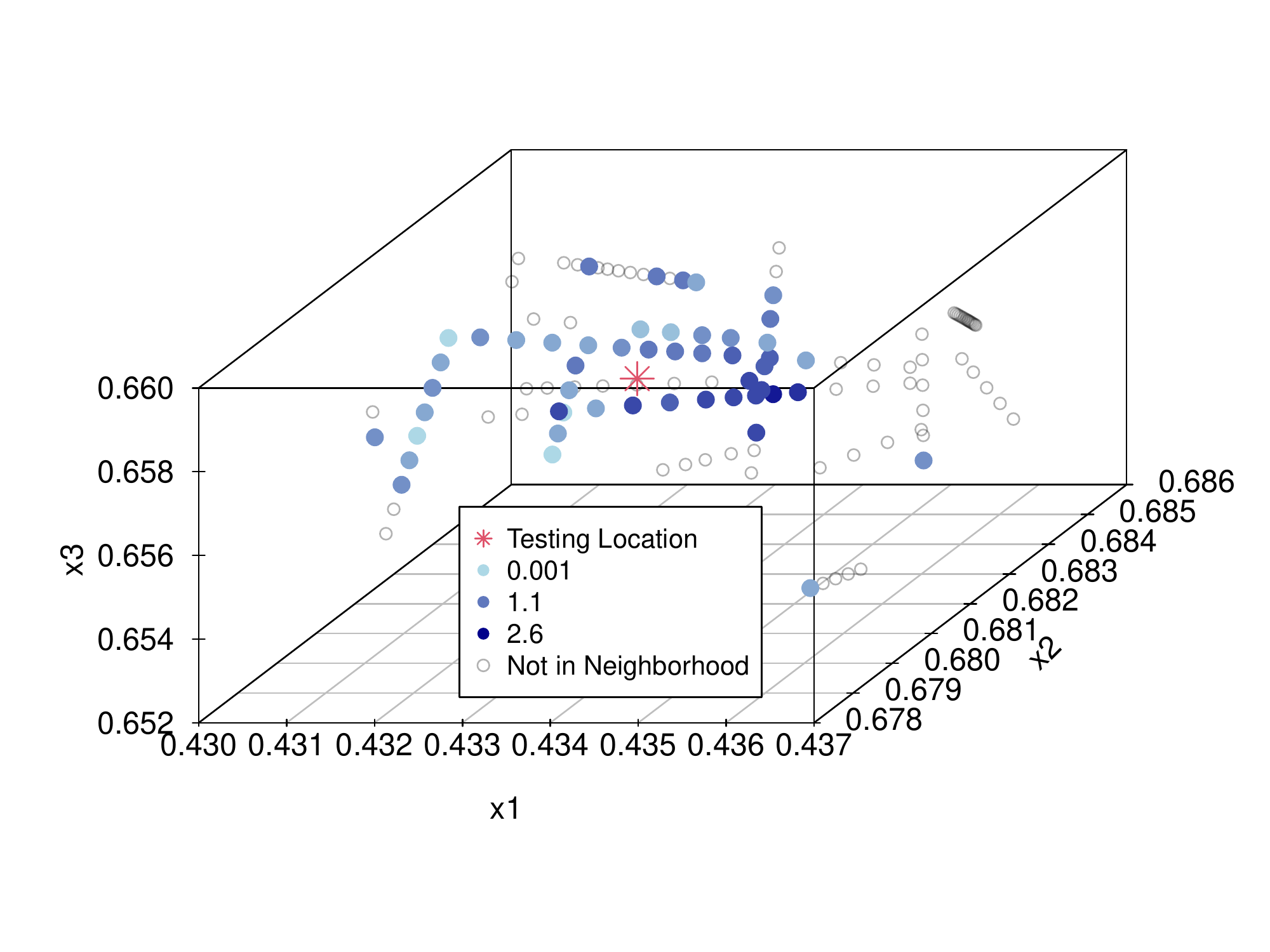}
	\caption{Left: Histogram of log LAGP nuggets with SVecchia and subset8k 
	nuggets. Right: An example neighborhood using LAGP's ALC for a point with 
	low estimated variance.}
	\label{fig:lagp_var}
\end{figure}

Even after such modifications, we found that LAGP and OK struggled to predict
at some testing sites.  {\tt GSLIB}, implementing OK, would simply refuse to
provide a prediction in these instances, or similarly when there are no
training data points within a user-specified radius (regardless of $m$),
returning an error code.  In these data, that amounts to about 300 testing
sites per fold.  The {\tt laGP} software would furnish a prediction, but when
comparing the corpus of other predictions in a fold it was obvious that
something was amiss, particularly with the estimated (local) nugget parameter
and, consequently, the predicted variance. To investigate, we plotted a
histogram of the estimated nuggets from all of the local fits, shown in the
left panel of Figure \ref{fig:lagp_var}, and compared these against the global
nugget(s) provided by subset and SVecchia methods.  We observed that
occasionally, local nuggets were being estimated at the lower-threshold
imposed by the {\tt laGP} default search range (leftmost-bin in the
histogram).  The local neighborhood for one, representative
member of this group is shown on the right panel of the figure.   Observe
that most of the points nearby have log gold values above 1 (are
similar shades of blue) with some satellite points having lower log gold
values. Thus a prediction of log gold a bit higher than 1 with low variance
makes sense. However, the true log gold value for this point is about 0.1,
meaning the prediction is confident and wrong -- which would lead to a
poor score evaluation.

Of course, we cannot know this location provides a bad prediction until we look 
at the testing value for this predictive location. So we decided to replace 
local nuggets estimated at the lower-bound of {\tt laGP}'s' search range with the 
median of the nuggets from the rest of the distribution.  This led to a 
substantial improvement in out-of-sample scores, described momentarily.  If 
this sounds {\em ad hoc}, that is because it is.  But a prediction with UQ that 
is based on compromise and a limited degree of post-hoc human intervention is 
better than no prediction at all (OK/{\tt GSLIB}).

\begin{figure}[ht!]
	\centering
	\includegraphics[width=5.6cm, trim=20 20 0 50]{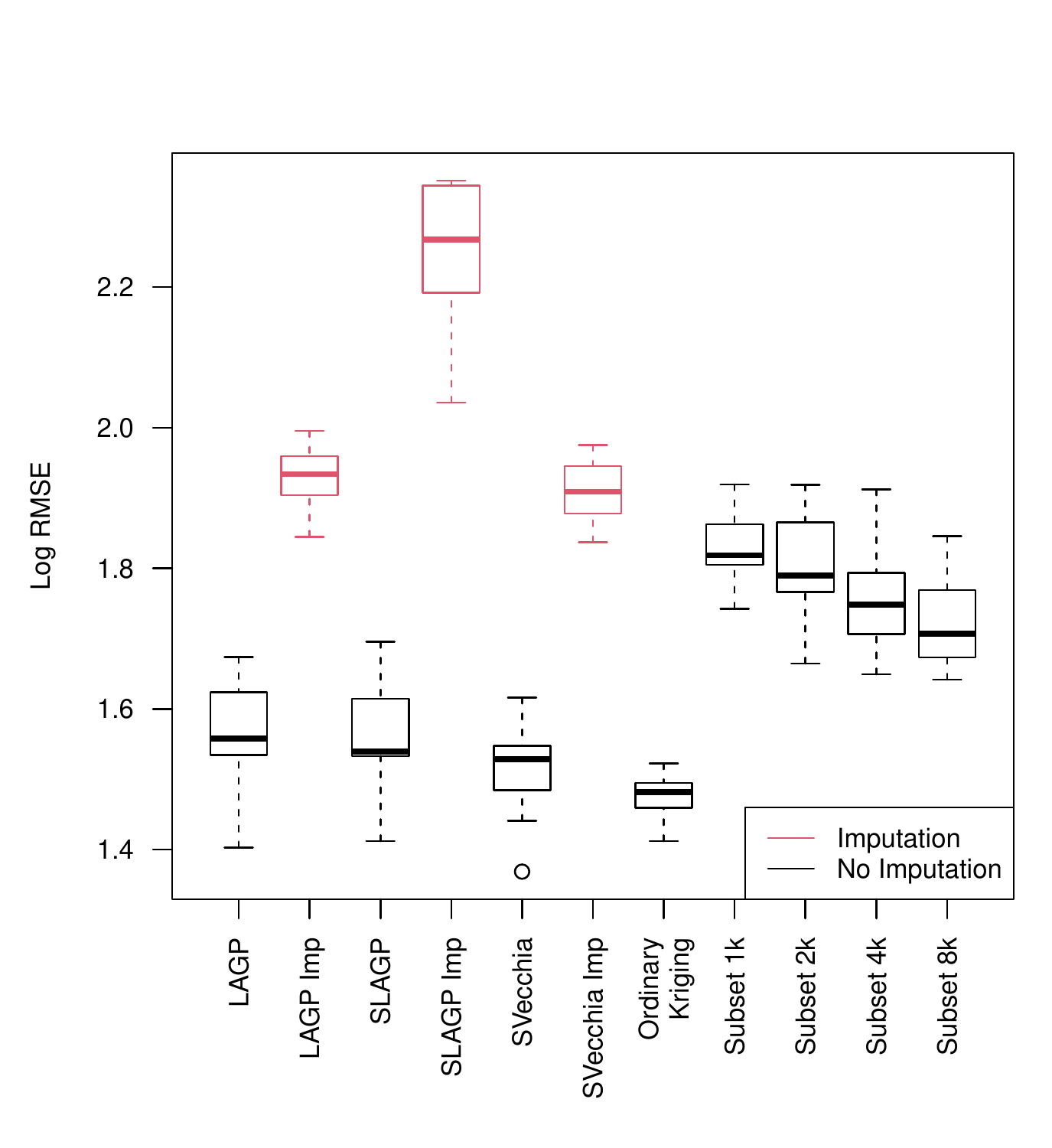}
	\includegraphics[width=5.85cm, trim=0 20 0 50]{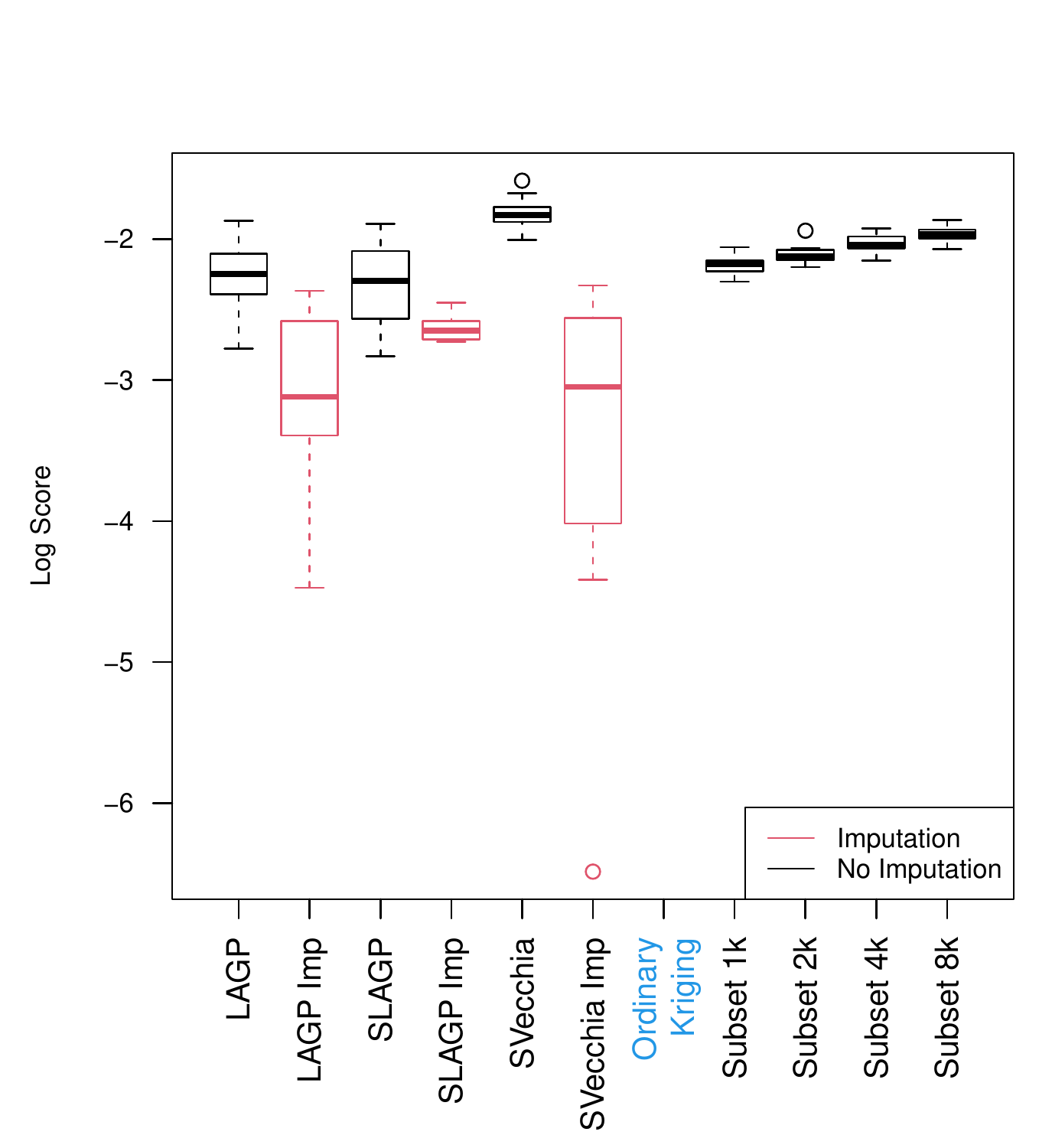}
	\includegraphics[width=5.7cm, trim=0 20 10 50]{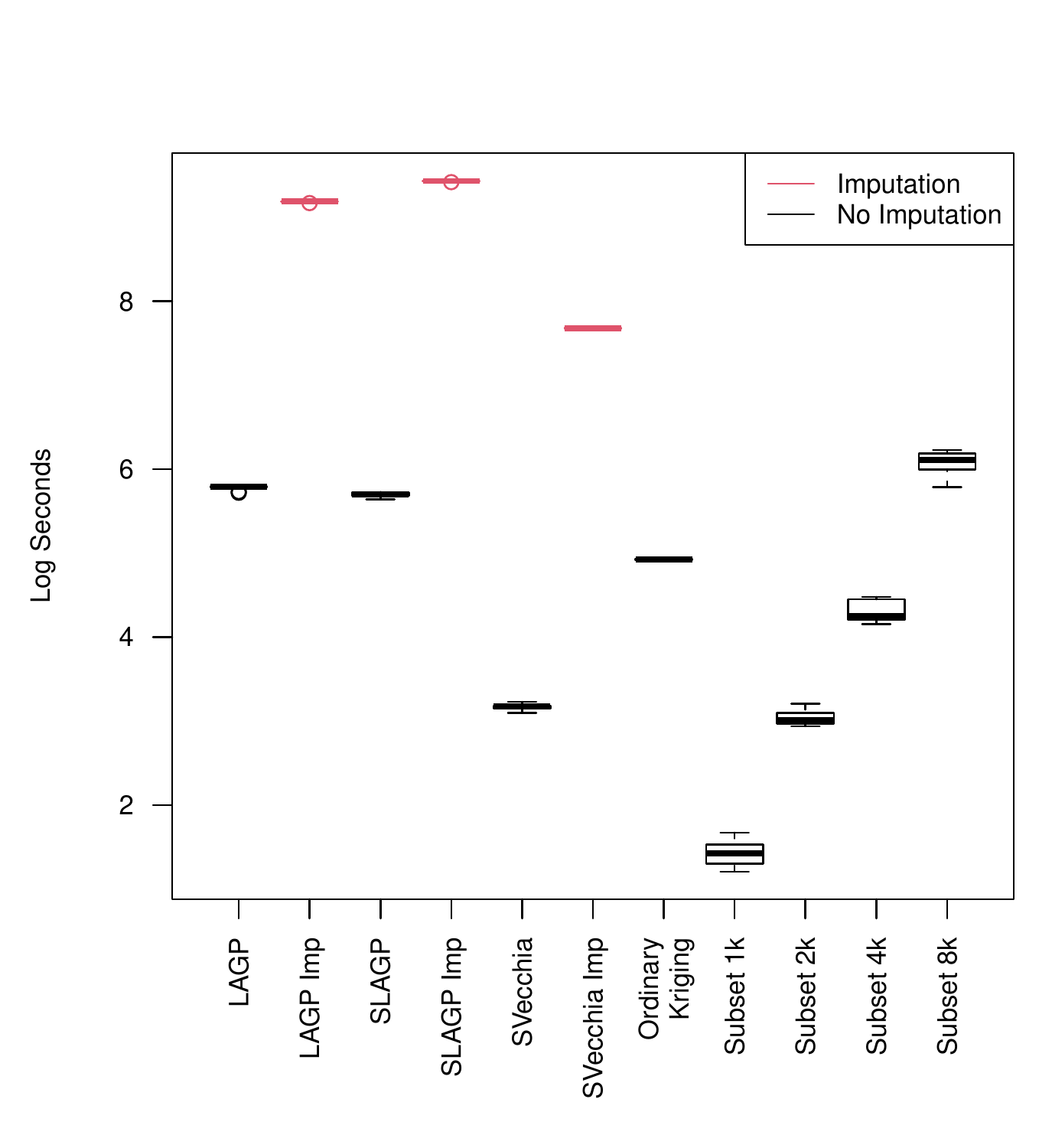}
	\caption{Drillhole-preserving 10-fold cross validation summary for the 
	first data set. See Figure \ref{fig:AGA} caption.  Red boxplots are 
	discussion in Section \ref{sec:imp}.}
	\label{fig:Skeena}
\end{figure}

Figure \ref{fig:Skeena} shows these results with black boxplots.  (The red
ones involve a study on imputation in Section \ref{sec:imp}, so ignore those
for now.) The story is similar to the first data set in Figure \ref{fig:AGA}:
SLAGP, SVecchia, and OK outperform subsetting in terms of RMSE.  Here, OK
appears to be the most accurate, but these RMSE calculations do not include
any error-coded outputs (representing more than 300 presumably ``bad''
predictions per fold).  So this is not a holistic assessment of OK accuracy.  By
score, which again cannot be calculated for OK, SVecchia is the clear winner, 
and the second-fastest in this comparison. (S)LAGP, which is similar in spirit 
to OK, has inferior scores despite modifications to address stability issues to
do with the nugget (above).

\subsection{Imputation}
\label{sec:imp}


It is unsatisfying to discard data.  Even a coarsely left-censored value
contains information, which can be used to enhance training.  Perhaps even
more importantly, one may wonder how accurately those censored values may be
predicted, thereby increasing the resolution of those measurements, by
borrowing information from higher-accuracy (training) data measurements
nearby.  This is a standard enterprise in statistical learning when fully
probabilistic generative modeling, like the GP, is used.  There are many
options when handling ``missing data,'' of which censoring is one example
\citep{Schafer1997, Little2002}.

One way to incorporate censored ore values, without destroying smoothness or
stationarity assumptions underlying GP spatial models, is through
\emph{imputation} \citep[e.g.,][Chapter 5]{Little2002}.  In our context,
imputation basically means generating a plausible response $Y$-value for
censored locations that both respects the censored measurement, and the
smoothness of the underlying spatial field learned through other, completely
observed data.  Once generated, the imputed value may be treated as if it were
a completely observed value going forward, say for prediction. Of course,
treating an imputed value as observed ignores the uncertainty in the
imputation. \emph{Multiple imputation} acknowledges that uncertainty by
randomly imputing several possible values and performing inference based on
the corpus of those imputed values, e.g., through averaging.

Illustrating how this could work in our ore context requires some notational
scaffolding. Let $D_N = (D_{\mathrm{obs}}, D_{\mathrm{cens}})$ represent the 
partition of the complete data set into its fully observed and censored 
components, respectively.  For example, $D_{\mathrm{obs}} = (X_{\mathrm{obs}},
Y_{\mathrm{obs}})$, may be the portion of the second data set we were
working with in Section \ref{sec:cv}, and $D_{\mathrm{cens}} =
(X_{\mathrm{cens}}, Y_{\mathrm{cens}})$ was the part we (temporarily)
discarded. Imputed values $Y_{\mathrm{imp}} (X_{\mathrm{cens}})$,
may be used to augment $D_{\mathrm{obs}}$ to obtain
$D_{\mathrm{imp}} = (D_{\mathrm{obs}}, (X_{\mathrm{cens}}, Y_{\mathrm{imp}}))$ via truncated Gaussian simulation
\begin{equation}
	Y_{\mathrm{imp}} \sim 
	\mathcal{N}_{N_{\mathrm{cens}}}(\mu_{\mathrm{obs}}(X_{\mathrm{cens}}), 
	\Sigma_{\mathrm{obs}}(X_{\mathrm{cens}})) 
	\mathbb{I}_{\{Y_{\mathrm{imp}} \leq Y_{\mathrm{cens}}\}}, \label{eq:tmvn}
\end{equation}
where $\mathbb{I}_{\{Y_{\mathrm{imp}} \leq Y_{\mathrm{cens}}\}}$ is an 
indicator function, returning 1 if $Y_{\mathrm{imp}} \leq Y_{\mathrm{cens}}$ 
and 0 otherwise. Quantities $\mu_{\mathrm{obs}}(X_{\mathrm{cens}})$ and 
$\Sigma_{\mathrm{obs}}(X_{\mathrm{cens}})$ are the
predictive moments (\ref{eq:pred}) of a GP fit 
conditioned on $D_{\mathrm{obs}}$.

Although software exists to sample from a truncated MVN directly
\citep[e.g.,][]{tmvtnorm}, such as those in (\ref{eq:tmvn}), in practice it
can be difficult to generate a sufficient number of values below
$Y_{\mathrm{cens}}$ when $N_{\mathrm{cens}}$ is of modest size, for example in
the hundreds \citep{Li2015}. A more customized approach that acknowledges the
form of our (approximate/large-scale) spatial surrogates helps. In the
(S)LAGP context, we may use Eq.~(\ref{eq:vecchia}) to sample
$Y_{\mathrm{imp}}$ from the truncated MVN (\ref{eq:tmvn}) one at a time,
conditioning on the previously sampled imputed values and the observed data.
LAGP is designed to look at each location in the testing set independently
which makes this setup work. On the other hand, SVecchia is designed to give a
global model approximation, so doing a similar one-at-a-time conditional
imputation is too crude. We instead prefer a bespoke rejection sampling 
\citep{Casella2004} scheme that proceeds in epochs: first generate posterior 
samples from the MVN (\ref{eq:tmvn}) unconstrained, keeping any values that 
satisfy the censoring threshold. Then condition on those imputations and the 
observed values, resampling at locations without an imputed value, repeating 
until $Y_{\mathrm{imp}}$ is completely filled in.

\begin{algorithm}[ht!]
	\caption{Multiple imputation for large scale GPs under left censoring }	
	\textbf{input} $D_{\mathrm{obs}}$ and 
	$D_{\mathrm{cens}}$ and testing locations $\mathcal{X}$\\
	\DontPrintSemicolon
	\For{i = $1, \dots, M$} {
		\setlength\parindent{24pt}
		\If{(S)LAGP} {
			$D = D_{\mathrm{obs}}$\;
			\For{j = $1, \dots, N_{\mathrm{cens}}$} {
				$(\mu, \sigma^2) = 
				\text{(S)LAGP}(D, X_{\mathrm{cens}}[j])$
				\tcp*{LAGP prediction: Sec.~\ref{sec:transductive}}
				$Y_{\mathrm{imp}}[j] \sim 
				N_{1}\left(\mu, \sigma^{2}\right) 
				\mathbb{I}_{\left\{Y_{\mathrm{imp}}[j] \leq 
				Y_{\mathrm{cens}}\right\}}$
				\tcp*{Posterior tnorm draw: Eq.~(\ref{eq:tmvn})}
				$D = \left(D, (X_{\mathrm{cens}}[j], 
				Y_{\mathrm{imp}}[j])\right)$
				\tcp*{Imputation step}
			}
			$(\mu_i, \sigma^2_i) = \text{(S)LAGP}(D, \mathcal{X})$
			\tcp*{Predict at testing locations}
		}
		\If{SVecchia} {
			$D = D_{\mathrm{obs}}$\;
			\While{$\vert X_{\mathrm{cens}}\vert > 0$} {
				$(\mu, \Sigma) = \text{SVecchia}(D, X_{\mathrm{cens}})$
				\tcp*{SVecchia prediction: Eq.~(\ref{eq:svecchiapred})}
				$Y_{\mathrm{imp}} \sim N_{\vert X_{\mathrm{cens}} 
				\vert} (\mu, \Sigma)$
				\tcp*{Unconstrained posterior MVN draw}
				$w = \text{which}(Y_{\mathrm{samp}} < Y_{\mathrm{cens}})$\;
				$D = (D, (X_{\mathrm{cens}}[w], Y_{\mathrm{imp}}[w]))$
				\tcp*{Imputation step}
				$X_{\mathrm{cens}} = X_{\mathrm{cens}} \setminus 
				X_{\mathrm{cens}}[w]$\;
			}
			$(\mu_i, \Sigma_i) = \text{SVecchia}(D, \mathcal{X})$			
			\tcp*{Predict at testing locations}
		}
	}
	\Return $(\mu_i, \Sigma_i)$, \text{for } $i_,\dots,m$
		\tcp*{where $\Sigma_i = \mathrm{diag}(\sigma_i^2)$ for (S)LAGP}
	\label{alg:multimp}
\end{algorithm}

Algorithm \ref{alg:multimp} provides pseudo-code for concreteness, wrapping a
single imputation with a {\tt for} to obtain $M$ multiple imputations.
\citet{Rubin1987} indicates that $M$ between two and ten works well, so we use
$M = 5$ in our exercises. Each of the $M$ imputed values are plausible
realizations of the censored measurements, which correspond to $M$
posterior/predictive Gaussian distributions for each testing location. Thus
we may use Gaussian mixture moment equations \citep{Reynolds2009} to report the
mean and variance predictions for the testing set:
\begin{align}
	\label{eq:mixpred}
	\mu_{\mathrm{MI}}(\mathcal{X}) &= \frac{1}{M} \sum_{i = 1}^M \mu_i 
	(\mathcal{X}) \\
	\sigma^2_{\mathrm{MI}} (\mathcal{X}) &= \frac{1}{M} \sum_{i = 1}^M 
	\sigma_i^2 (\mathcal{X}) + \frac{1}{M} \sum_{i = 1}^M \mu^2_i  
	(\mathcal{X}) - \left(\frac{1}{M} \sum_{i = 1}^M \mu_i 
	(\mathcal{X})\right)^2 \nonumber
\end{align}
While this cannot completely account for all possible uncertainties due to
imputation, because we have not looked at all possible imputation values (only
$M$), we can always increase $M$ if desired. It may be shown that
these equations give an unbiased estimate of mean and variance for any $M$.


\subsubsection*{Imputation in practice}

\begin{figure}[ht!]
	\centering
	\includegraphics[height=7cm,trim=0 20 20 50]{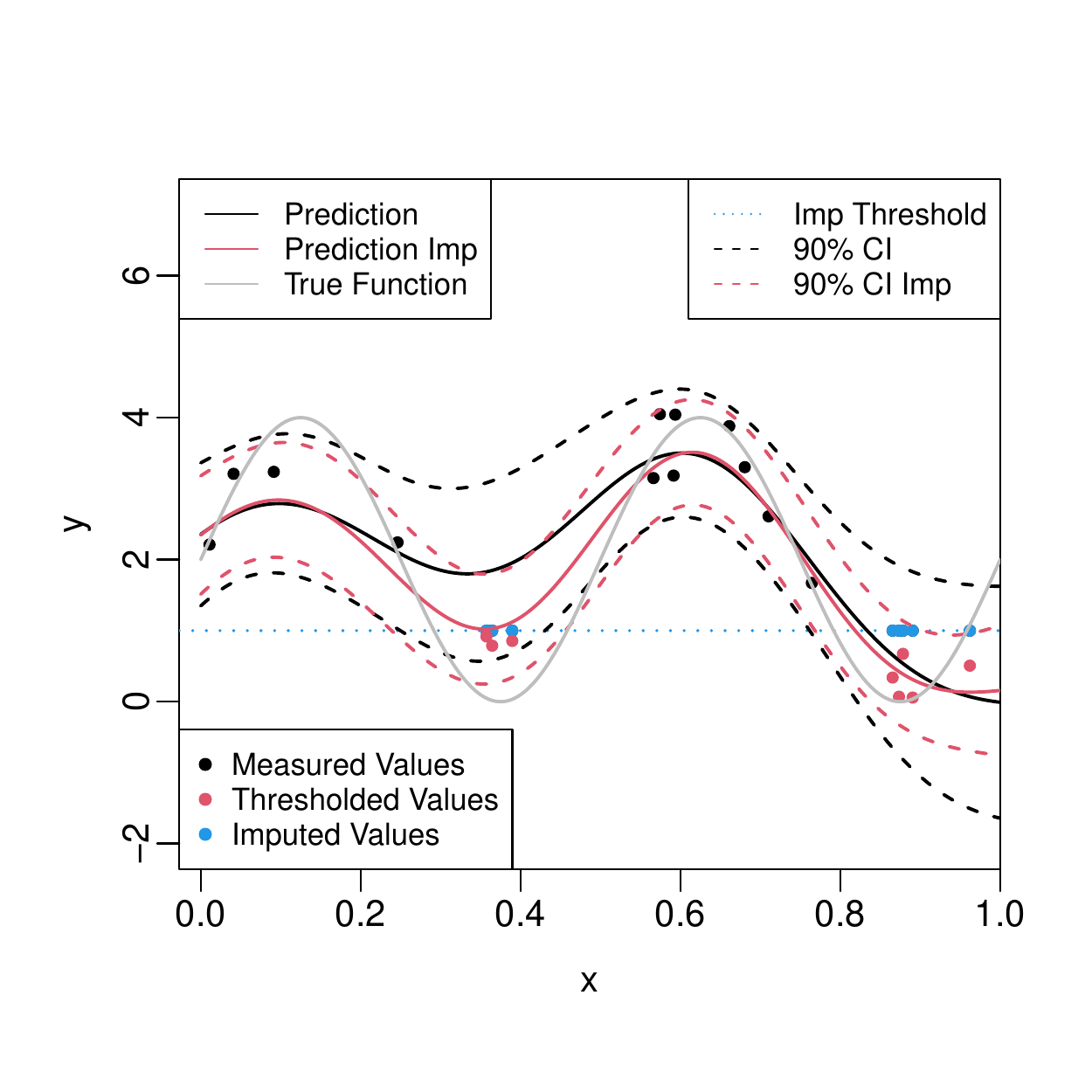}
	\includegraphics[height=6.9cm,trim=0 20 20 50]{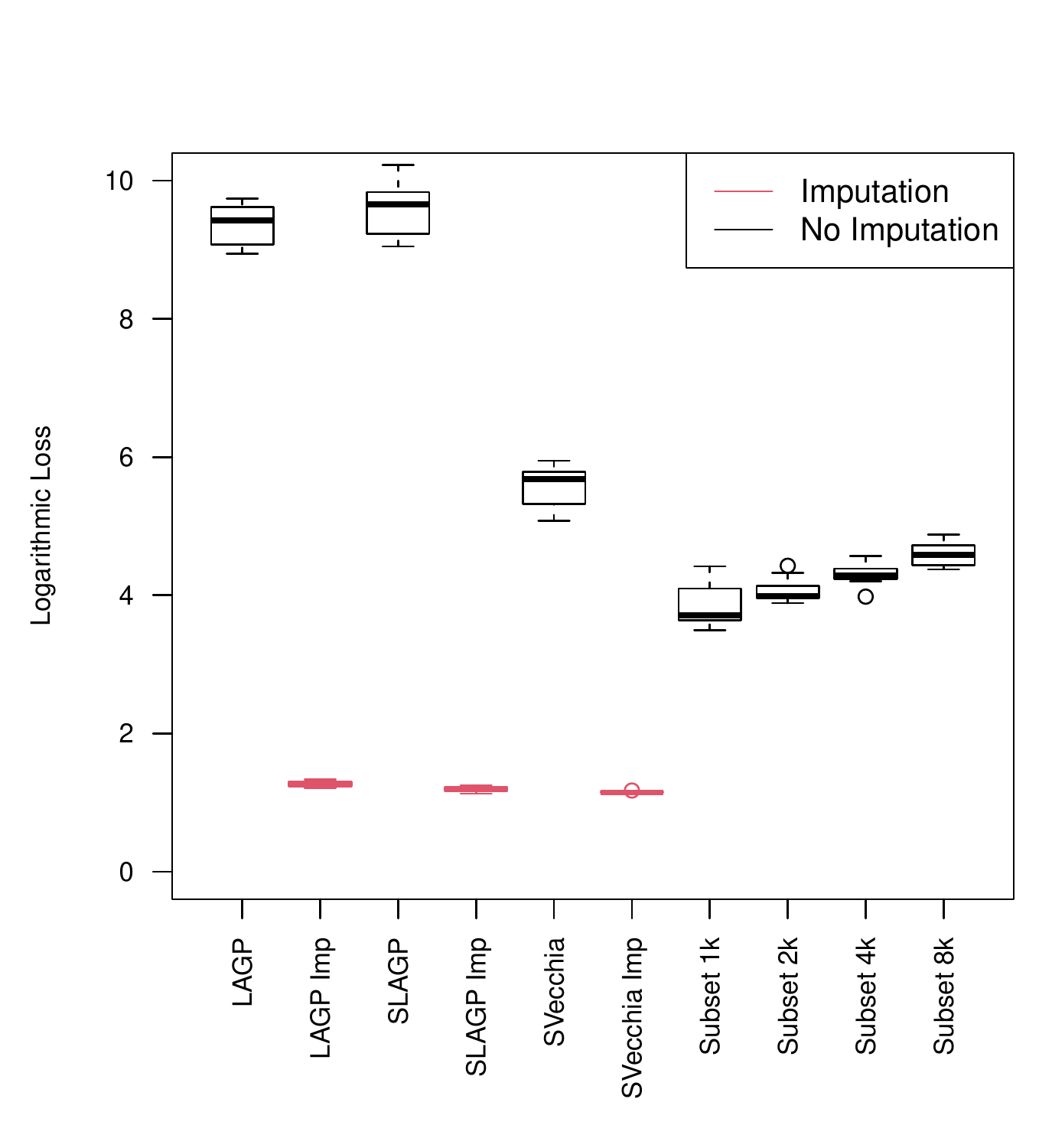}
	\caption{Left: Imputation illustration for 1d synthetic data. Right: CV
	log loss (\ref{eq:logloss}) from ore data set
	two that are below the detection threshold using big data methods with and
	without imputation.}
	\label{fig:Imputation}
\end{figure}

To begin with an illustration in a simple, controlled setting, the left panel
of Figure \ref{fig:Imputation} shows a classical GP model with and without
imputation. The true function is $f(x) = 2\text{sin}(4\pi x) + \epsilon$ where
$\epsilon
\sim \mathcal{N}(0, 0.1)$ and an observation threshold of $y = 1$ with $n=20$
randomly selected training points. There are two main regions of censoring,
one in the center and one at the upper end of inputs. In the center, the model
with imputation gets closer in mean to the truth.  On the upper end, the
variance of our predictions is much lower when conditioning on imputed values.
Observe that the model with imputation is a
better fit for the true curve. 

When working with real data, we do not have the luxury of knowing what the
``true curve'' is.   Consequently, it is much tougher to assess improvement in
the quality of fit through imputation.  RMSE and score on an out-of-sample
testing set, say via CV, are problematic because censored data values -- as
elements of the testing set -- are not realizations from the same population
as the model/predictive quantities are ($\mu_n(x)$ and $\sigma^2_n(x)$).  The
only ``truth'' we know about these points is that they are measured to be
below the threshold. The best we can to is determine if the model accurately
predicted the ``parity'' of its recorded value, either above or below the
threshold. The GP framework makes this transition simple. The
probability of accurate prediction under a threshold may be obtained through
inverse Gaussian CDF evaluated at the threshold. The proper scoring mechanism
\citep{Gneiting2007} for such probabilities is the logarithmic loss
\citep[LL;][]{Good1952}, also known as cross-entropy loss in the neural
network literature. Lower LL is better.  When all of the testing data are from
one class (less than $Y_{\mathrm{cens}}$), LL boils down to:
\begin{equation} \label{eq:logloss}
	\mathrm{LL}(\mathcal{X}_{\mathrm{cens}}) = 
	-\frac{1}{N'_{\mathrm{cens}}}\sum_{x \in 
		\mathcal{X}_{\mathrm{cens}}}  \log(\mathbb{P}(Y(x) \leq Y_{\mathrm{cens}}) 
		= 
	-\frac{1}{N'_{\mathrm{cens}}}\sum_{x \in \mathcal{X}_{\mathrm{cens}}} 
	\log\left(\Phi^{-1}(Y_{\mathrm{cens}}; \mu(x), \sigma^2(x))\right). 
\end{equation}

Returning now to our second ore analysis from Section \ref{sec:cv}, we
describe our experience with multiple imputation on these data. We only
explore (S)LAGP and SVecchia in this context. When dealing with the subset
methods, imputation is of limited additional value as completely observed data
are plentiful relative to the subset size.  It is cumbersome, but not
impossible to entertain imputation under OK. Being faithful to the imputation
scheme would require human intervention to re-fit variograms after each new
imputation is obtained.  That would result in over 100K variogram fits in this
example! In this context, we see LAGP as an equivalent, automatic variation on
OK that can be more easily entertained in an imputation/censored values
context.

The right plot of Figure \ref{fig:Imputation} demonstrates that our imputation
scheme yields improved LL (\ref{eq:logloss}) across the board. For reference,
$-\log(0.001) = 6.9$ and $-\log(0.2) = 1.6$, so fits leveraging imputation
provide a probability of 0.2 for being below thresholds, on average, compared
to 0.001 or worse for the models without imputation. 
At first, it is hard to square this improvement with what appears to be
contrary messaging from the imputation results in Figure \ref{fig:Skeena} (red
boxplots). If a practitioner is certain that a particular region is high in
gold concentration, {\em a priori}, then dropping the censored values (i.e.,
no imputation), which are all low-measurements, leads to more accurate
results. Yet dropping thresholded values exposes the practitioner to
confirmation bias: predictions appear more accurate in one part of the space at
the expense of massive over-predictions in another. This is what the right
panel of Figure \ref{fig:Imputation} shows.


\section{Conclusion} 
\label{sec:conclusion}

We have highlighted similarities and differences in GP modeling and kriging.
On the one hand, disparate terminology obscures a practically identical
modeling apparatus. On the other, approaches to hyperparameter estimation and
the treatment of anisotropy can imply drastically different degrees of
automation/requisite human intervention.  We demonstrated that with smaller
training data sets, expert intervention through variography can lead to
(slightly) better fits. However, we advocate for taking the human out of the
loop even in this situation, and relying on the likelihood.  We are not saying
the human should be eliminated entirely. Identifiability issues
\citep{Tang2021,Kaufman2013} inherent in separating signal from noise (i.e.,
inferring the nugget), and in determining smoothness ($p$ and $\nu$), mean it
is always sensible to inspect outputs and challenge downstream inferences
against stylized facts about the system.  But intimate human involvement
challenges reproducibility and limits scope for extension, such as with with
censored data in our mining context.  Careful engineering atop of sound
principles can improve upon many human labor-intensive tasks -- possibly a
defining feature of the machine learning approach to statistics.

By introducing modern kriging/GP approximations, we have shown that when
working with large mining datasets in low dimension, the scaled Vecchia
approximation is faster and at least as accurate as LAGP/OK with the benefit
of giving full joint posterior distributions.   Throughout, we leveraged
off-the-shelf libraries with defaults settings.  These represent the tip of an
iceberg.  See, e.g., \cite{Heaton2019} for a survey of additional modern
approaches to modeling spatial data. Mining data often have some
non-stationarity due to geological structures, e.g., faults, folds, and
fracture networks. Recently it has been suggested that a deep GP
\citep{Damianou2013, Sauer2022} may handle such non-stationary data more
gracefully than ordinary GPs: combining the accuracy of deep learning (deep
neural networks) with the uncertainty quantification (UQ) features of fully
probabilistic (GP) modeling.

Finally, GP-based machinery -- as opposed to kriging -- is less intimately
tied to low-dimensional (2-3d) input spaces, allowing for the incorporation of
more predictor variables. In the case of mining for gold, measurements for
other minerals (in a similar domain) can be used to help build models that are
more accurate and give better UQ. This can easily increase the input dimension
to the tens of variables.  That challenges variography, but the GP approach
remains essentially unchanged. 


\bigskip \noindent
{\bf Code Availability.}  All examples/illustrations can be found on our Git repo on Bitbucket:
\begin{center}
\url{https://bitbucket.org/gramacylab/mining_code/src/master/}
\end{center}

\bigskip \noindent
{\bf Data Availability.}  Data for the illustrative examples in Section
\ref{sec:review} is provided either in-line with our code (see git repo
above), or as included with an open source library utilized by our code. The
mining data from our industrial partners is proprietary and we do not have
permission to share it.  However, the data is in a standard CSV-column format,
and our code for those examples (in the repo above) may be utilized with any
data that can be supplied in that form.

\subsection*{Acknowledgements}
This work was conducted within the NSF I/UCRC Center for Advanced Subsurface 
Earth Resource Models (CASERM) which is an industry-university research center 
jointly managed by Colorado School of Mines and Virginia Tech under the NSF 
award numbers 1822108 and 1822146, respectively. The authors extend sincerest 
thanks to Alex Mason Apps at AngloGoldAshanti and Kelly Earle at Skeena 
Resources for supplying this project with mine assay data. This work was also 
supported by the U.S. Department of Energy, Office of Science, Office of 
Advanced Scientific Computing Research and Office of High Energy Physics, 
Scientific Discovery through Advanced Computing (SciDAC) program under Award 
Number 0000231018.

\subsection*{Declarations}
\textbf{Conflict of Interest.} The authors have no competing interests to 
declare.

\bibliographystyle{jasa}
\bibliography{references}

\begin{appendices}

\section{A common 2d kriging example}
\label{sec:meuse}

Consider the Meuse river 
data, which often serves as a tutorial-level example of kriging 
\citep{Pebesma2009}. These data consist of measurements 
of concentrations of metals in the topsoil of the Meuse river in the 
Netherlands. Our analysis here focuses on 155 zinc measurements in parts 
per million (ppm) taken at 2d spatial locations recorded as Rijksdriehoeks 	
Coordinates (National Triangulation Coordinates of the Neatherlands) 
distributed at varying uniformity along the river. To prepare the data for 
analysis, we follow the statistical learning/surrogate modeling practice of 
coding these inputs as $X_N$ in the unit cube $[0,1]^2$ and model the 
centered (i.e., empirically mean adjusted) natural logarithm of zinc 
concentration as the response. This
automatic pre-processing step helps initialize the numerical solvers for
hyperparameters at default settings, and makes a Gaussian assumption on the
response more compatible.
	
\begin{figure}[ht!]
	\begin{minipage}{11.2cm}
		\includegraphics[height=5cm,trim=15 30 20 50]{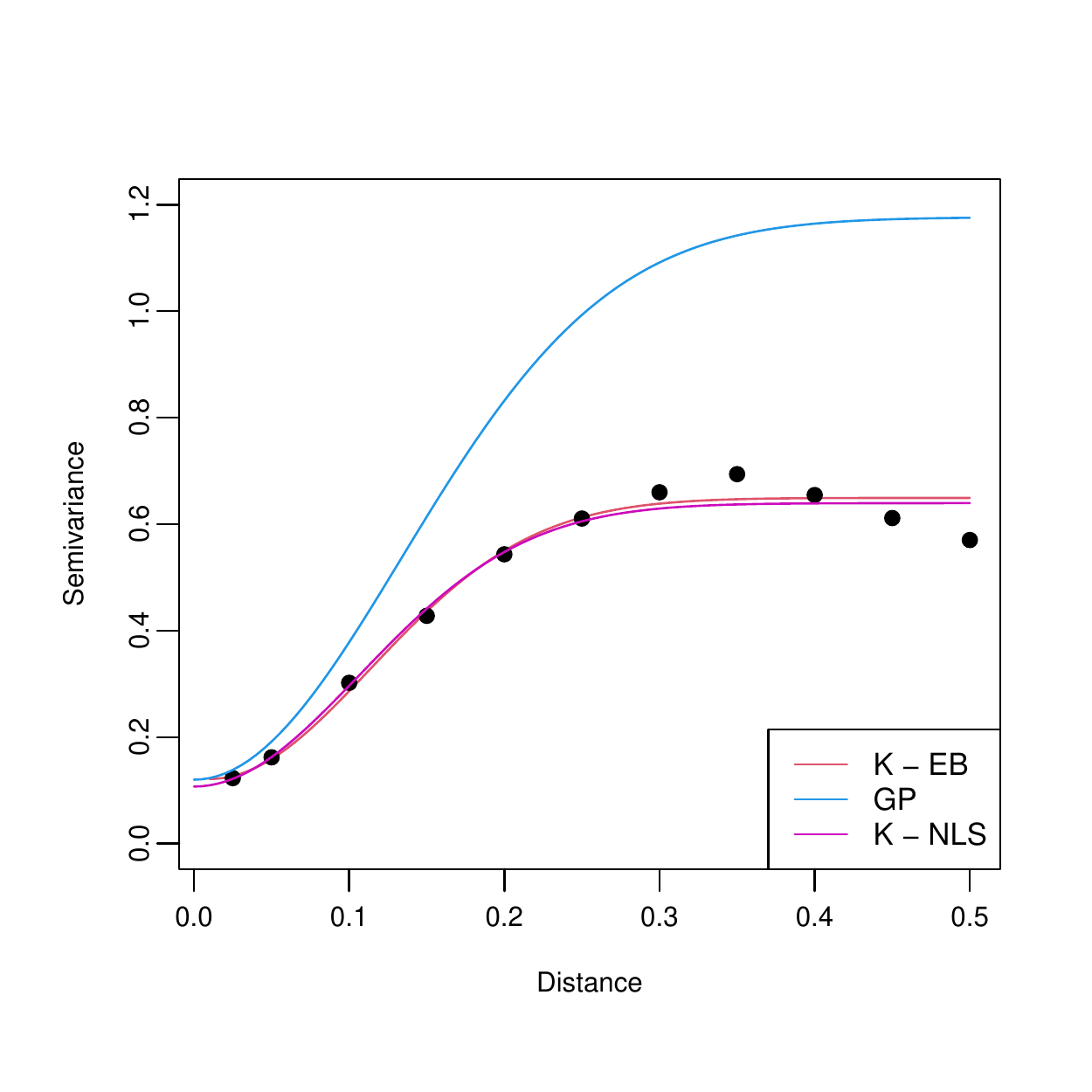}
		\includegraphics[height=5cm,trim=25 30 20 50]{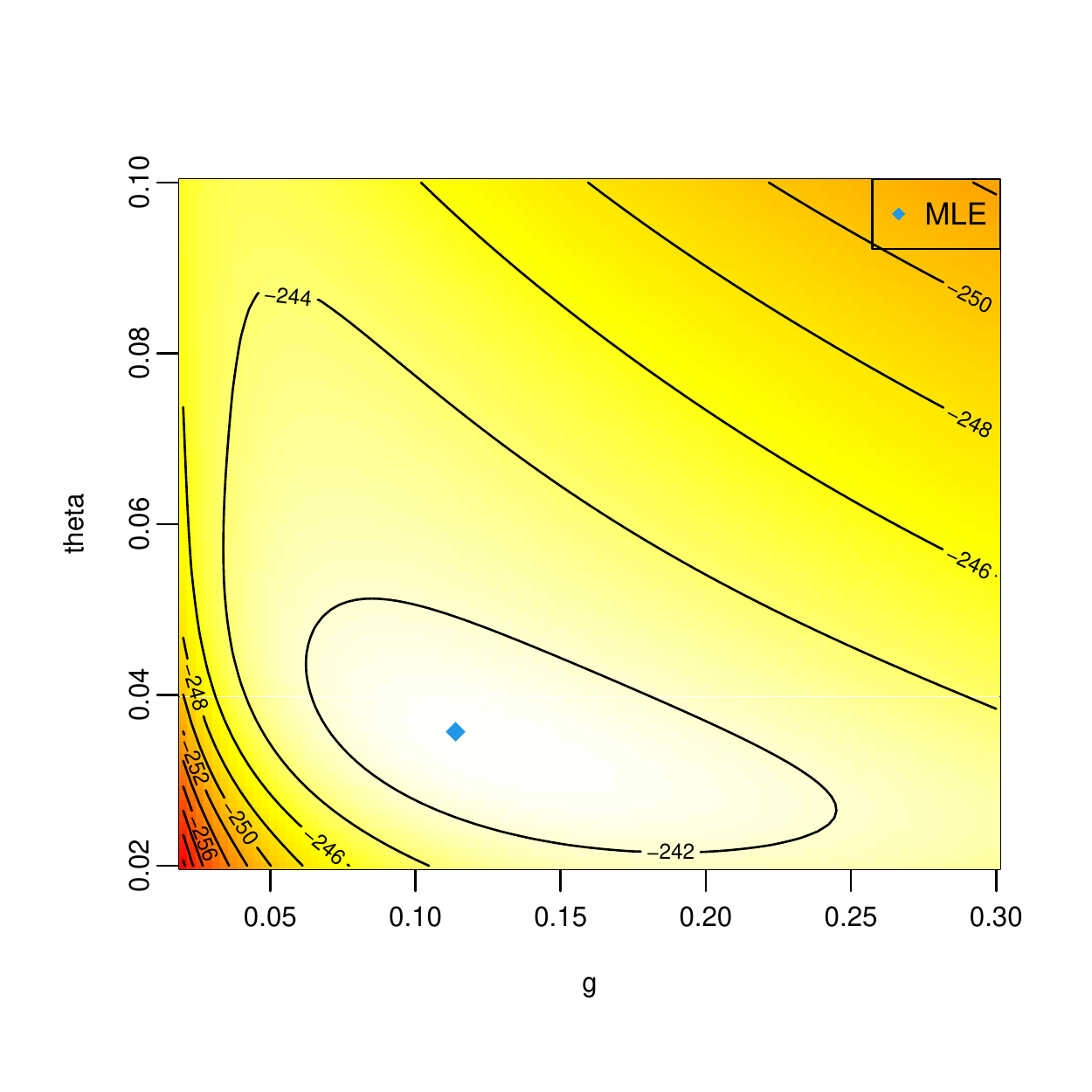}
	\end{minipage}
	\hfill
	\begin{minipage}{6cm} \footnotesize
		\begin{tabular}{l|l|l|l|}
			& GP & K-EB & K-NLS\\
			\hline
			$\tau^2$ or $\sigma^2$ & $1.06$ & $0.53$ & $0.53$\\
			\hline
			$\tau^2g$ or $\tau_k^2$ & $0.12$ & $0.12$ & $0.11$\\ 
			\hline
			$\theta$ or $R$ & $0.19^2$ & $0.15$ & $0.15$\\
			\hline
		\end{tabular}
	\end{minipage}
	\caption{Left: Variogram for kriging by eye in red, kriging by NLS in 
	purple and GP in blue. Middle: GP log MLE surface with optimum in 
	blue. Right: Table comparing hyperparameters.}
	\label{fig:meuse_inf}
\end{figure}
	
\begin{figure}[ht!]
	\includegraphics[height=6cm,trim=20 30 20 50]{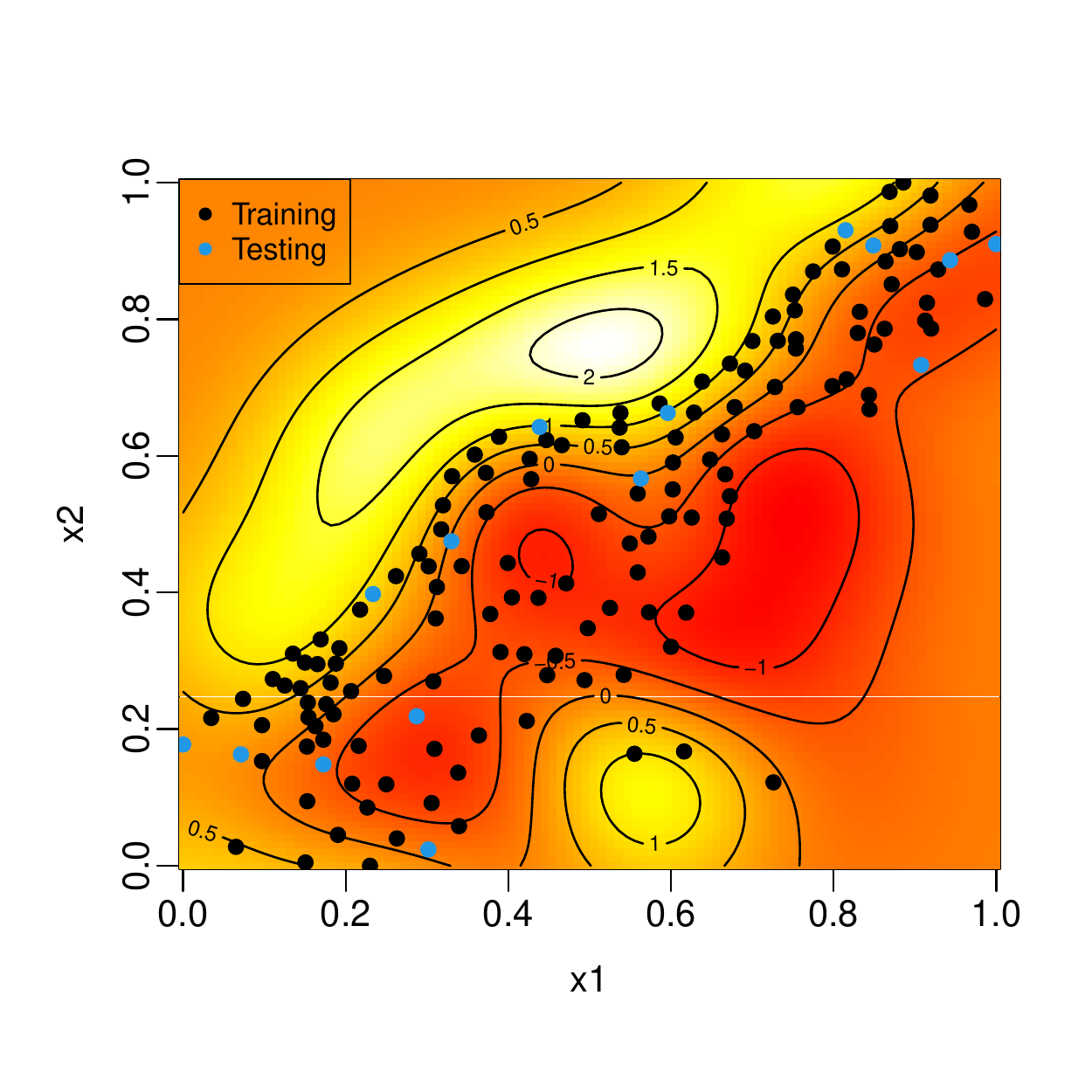}
	\includegraphics[height=6cm,trim=20 30 20 50]{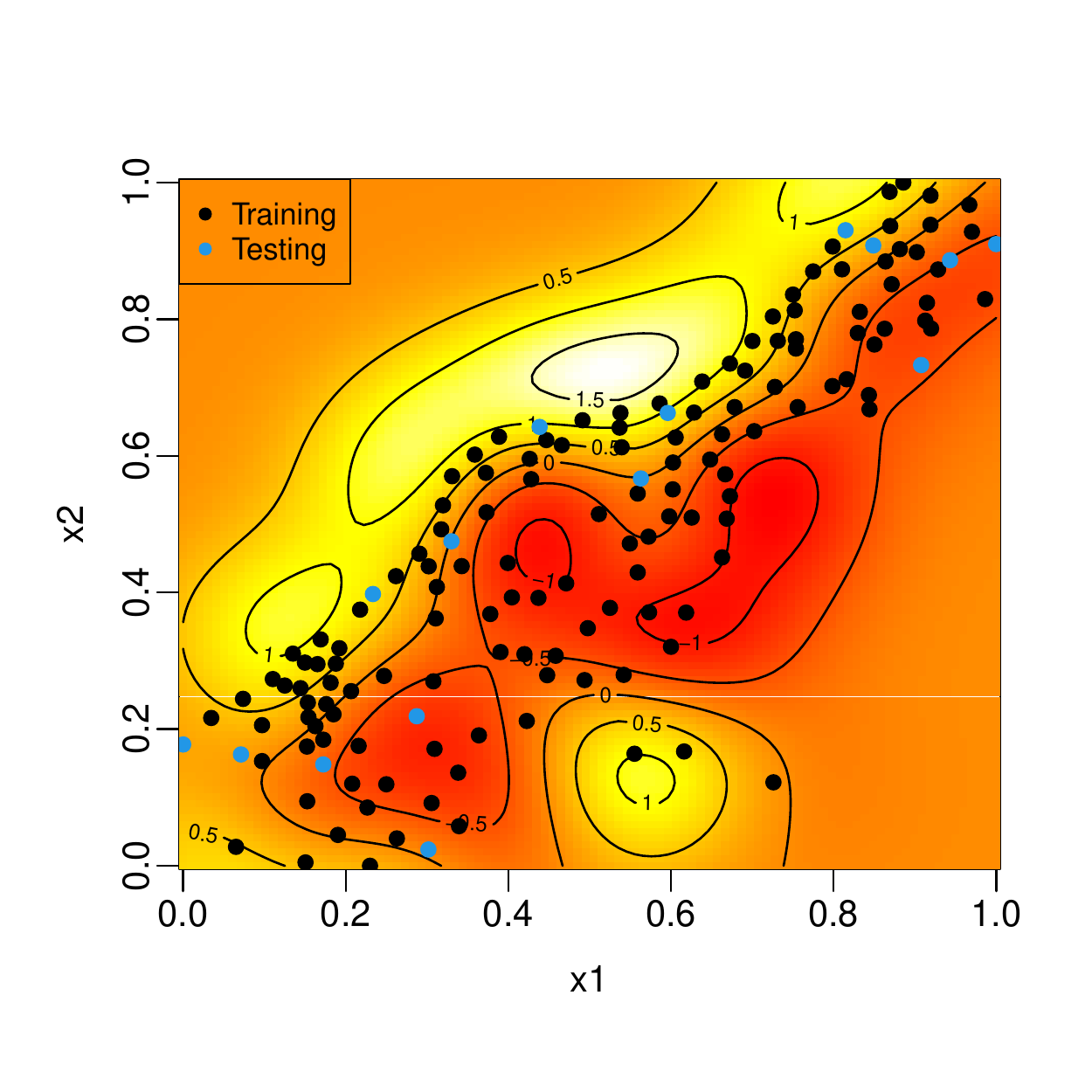}
	\caption{GP-MLE (left) and K-EB (right) predictions of log zinc.}
	\label{fig:meusepred}
\end{figure}

Figure \ref{fig:meuse_inf} shows the variogram and MLE-based analysis in views
similar to Figures \ref{fig:loglik} and \ref{fig:Vgram} for these data.
Observe that the empirical semivariogram is much better behaved, compared to
our first example, and that consequently EB and NLS hyperparameter estimates
and variogram fits are quite similar.  The main difference between the GP and
these kriging alternatives is that the estimated scale hyperparameter $\tau^2$
is much larger.  With similar nuggets and lengthscales, the rate of increase
in the variogram is about the same, but reaching a higher level.  Since the
estimated scale does not impact the predictive mean $\mu(\mathcal{X})$, see
Eq.~(\ref{eq:pred}), it is perhaps not surprising that the predictive surfaces
(via $\mathcal{X}$ comprised of a $100 \times 100$ grid)
for GP-MLE and K-EB are so similar in Figure \ref{fig:meusepred}.
A surface for K-NLS is not shown because of the stark similarity with K-EB.
Unfortunately, we do not know the real response(s) on this predictive grid, so
an out-of-sample analysis must be more limited than our first example. As a
pre-cursor to the cross-validation scheme introduced in Section
\ref{sec:results}, we consider a simple random 90:10 train--test partition of
our original 155 training data records.  This results in $N=140$ values
$(X_N,Y_N)$ that can be used for training our three models, and $N'=15$ 
values $(\mathcal{X},Y(\mathcal{X}))$ quarantined away for testing to see which 
works best out-of-sample by RMSE and score (again $\mathrm{score}_f$ from
Eq.(\ref{eq:rmse})).  Observe that in this case there is no distinction
between ``true'' and random $y$-values as we only have the testing outputs
from the data.  In Figure \ref{fig:meusepred}, the random partition is
indicated by coloring the 15 testing dots blue as opposed to black.  
\begin{table}
	\begin{tabular}{l|l|l|l|}
		Meuse & GP & K-EB & K-NLS\\
		\hline
		RMSE & 0.103 & 0.109 & 0.108\\
		\hline
		score$_\mathrm{f}$ & 16.79 & 15.97 & 16.50\\ 
		\hline
	\end{tabular}
	\caption{RMSE and score for GP and kriging on the Meuse river 
	data on a set of 15 hold out points.}
	\label{tab:meusescores}
\end{table}
Table \ref{tab:meusescores} records these values. Notice that while GPs edge out 
both kriging methods, the results are comparable for all three. Ideally we 
would repeat this enterprise multiple times, with novel 90:10 splits.  For the
MLE-GP this is easy, but for the kriging alternative it can be labor 
intensive if eyeballs are involved.

\end{appendices}

\end{document}